\documentclass[10pt]{wlscirep}
\usepackage[utf8]{inputenc}
\usepackage[T1]{fontenc}
\usepackage[inkscapelatex=false]{svg}
\usepackage{subcaption}
\usepackage{amsmath}
\usepackage[toc,page]{appendix}
\usepackage{tabularx}
\usepackage{newfloat}
\usepackage{makecell}
\DeclareFloatingEnvironment[name={Supporting Figure}]{supportingfigure}
\usepackage{url}

\DeclareFloatingEnvironment[name={Supporting Table}]{supportingtable}

\definecolor{red}{rgb}{0,0,0}

\title{Disentangling the Complex Multiplexed DIA Spectra in De Novo Peptide Sequencing}

\author[1,+]{Zheng Ma}
\author[1,+]{Zeping Mao}
\author[1]{Ruixue Zhang}
\author[2]{Jiazhen Chen}
\author[3]{Lei Xin}
\author[3]{Baozhen Shan}
\author[2,*]{Ali Ghodsi}
\author[1,*]{Ming Li}
\affil[1]{Cheriton School of Computer Science, University of Waterloo, Waterloo, N2L 3G1, Canada}
\affil[2]{Department of Statistical and Actuarial Science, University of Waterloo, Waterloo, N2L 3G1, Canada}
\affil[3]{Bioinformatics Solutions Inc., Waterloo, ON, N2L 3K8, Canada}

\affil[*]{corresponding authors, ali.ghodsi.uwaterloo.ca mli@uwaterloo.ca}
\affil[+]{these authors contributed equally to this work}

\begin{abstract}

Data-Independent Acquisition (DIA) was introduced to improve sensitivity to cover all peptides in a range rather than only sampling high-intensity peaks as in Data-Dependent Acquisition (DDA) mass spectrometry. However, it is not very clear how useful DIA data is for de novo peptide sequencing as the DIA data are marred with coeluted peptides, high noises, and varying data quality. We present a new deep learning method DIANovo, and address each of these difficulties, and improves the previous established systems by a large margin, via equipping the model with a deeper understanding of coeluted DIA spectra.  This paper also provides criteria about when DIA data could be used for de novo peptide sequencing and when not to by providing a comparison between DDA and DIA, in both de novo and database search mode. We find that while DIA excels with narrow isolation windows on older-generation instruments, it loses its advantage with wider windows. However, with Orbitrap Astral, DIA consistently outperforms DDA due to narrow window mode enabled. We also provide a theoretical explanation of this phenomenon, emphasizing the critical role of the signal-to-noise profile in the successful application of de novo sequencing.
\end{abstract}

\begin{document}

\flushbottom
\maketitle
%
%
\thispagestyle{empty}




\section{Introduction}

De novo peptide sequencing plays a crucial role in proteomics, enabling the identification of novel peptides, post-translational modifications, and mutations absent from existing protein databases\cite{lu2004denovo}\cite{tran2017deepnovo}\cite{frank2005pepnovo}\cite{zhang2012denovo}. This capability is crucial for personalized immunotherapy, guiding targeted treatments by identifying unique neoantigens. It is also essential for studying species with unsequenced genomes, where database searches are not feasible~\cite{fernandez1998dbsearch}~\cite{bassani2016neoantigen}~\cite{vitiello2017neoantigen}.

Compared to database-driven methods, de novo sequencing offers the advantage of discovering new peptide sequences independently of pre-existing data, enables capturing the full diversity of proteomes, especially in complex and dynamic biological systems~\cite{vitiello2017neoantigen} \cite{anonymous2017problemneoantigen} \cite{bassani2016direct}. In traditional Data-Dependent Acquisition (DDA) methods, where abundant peptides are selected to fragment\cite{yates1995dda} \cite{michalski2011ultra}, deep learning-based approaches such as DeepNovo\cite{tran2017deepnovo}, Casanovo\cite{yilmaz2022casanovo}, PepNet\cite{liu2023accurate}, and GraphNovo\cite{mao2023graphnovo} have significantly improved performance, making them valuable and efficient tools for biological research.

Despite their promising results, traditional DDA methods suffer from limitations such as biased sampling and inconsistent detection of low-abundance peptides, leading to incomplete proteome coverage~\cite{michalski2011ultra}~\cite{venable2004diaintro}~\cite {gillet2012targeted}. Data-Independent Acquisition (DIA) \cite{gillet2012targeted}\cite{li2021dia} addresses these limitations by fragmenting all ionized peptides within a predefined mass range, providing a more comprehensive and unbiased snapshot of the proteome~\cite{venable2004diaintro}.

Data-independent acquisition (DIA) presents unique challenges for peptide sequencing compared to data-dependent acquisition (DDA). Unlike DDA, which typically produces spectra associated with a single precursor ion, DIA generates highly multiplexed spectra in which fragment ions from multiple coeluting peptides coexist \cite{rost2014openswath} \cite{egertson2015multiplexed}. This concurrent fragmentation increases spectral complexity and introduces substantial noise, making it difficult to distinguish true signals from background interference. The overlap of fragmentation patterns further complicates fragment-ion assignment, thereby posing significant obstacles for de novo peptide sequencing and requiring more computationally intensive analysis pipelines. Despite these challenges, DIA also provides rich chromatographic information by capturing the temporal profiles of peptide ions. These chromatogram traces, when compared across coeluting peptides, can offer valuable clues for linking fragment ions to their peptide sources, thus partially mitigating the interpretative difficulties inherent in DIA data and offering advantages for downstream peptide identification and quantification.

Recent advances such as DeepNovo-DIA~\cite{tran2019deepnovodia}, Transformer-DIA~\cite{ebrahimi2024transformerbasednovopeptidesequencing}, and Cascadia~\cite{sanders2024cascadia} have leveraged deep learning to address the unique challenges of de novo sequencing in DIA. These models employ neural architectures to jointly analyze precursor and fragment ions across multiple dimensions, including mass-to-charge ratio ($m/z$), retention time, and intensity, thereby enabling more effective interpretation of highly multiplexed spectra. For example, DeepNovo-DIA integrates Ion-CNN and Spectrum-CNN~\cite{lecun1998cnn} with a long short-term memory (LSTM) network~\cite{hochreiter1997lstm} to capture both the three-dimensional structure of fragment ions and their correlations. Transformer-DIA, in contrast, adopts a Transformer-based encoder–decoder framework~\cite{vaswani2017attention}, autoregressively predicting peptide sequences by generating each amino acid conditioned on the previously decoded subsequence. Cascadia~\cite{sanders2024cascadia} further extends this paradigm by adapting a Transformer model originally designed for DDA data to DIA through the use of large-scale training data and spectrum augmentation strategies. Despite these successes, current approaches remain limited by their tendency to encode only subsets of the available spectral evidence, often neglecting informative signals arising from coeluting peptides, which may reduce sensitivity for detecting challenging peptide sequences.

\begin{figure}[!ht]
\centering
\begin{minipage}{0.76\textwidth} 
  \centering

  \hfill
  \begin{subfigure}[b]{0.63\textwidth}
    \includegraphics[width=\textwidth]{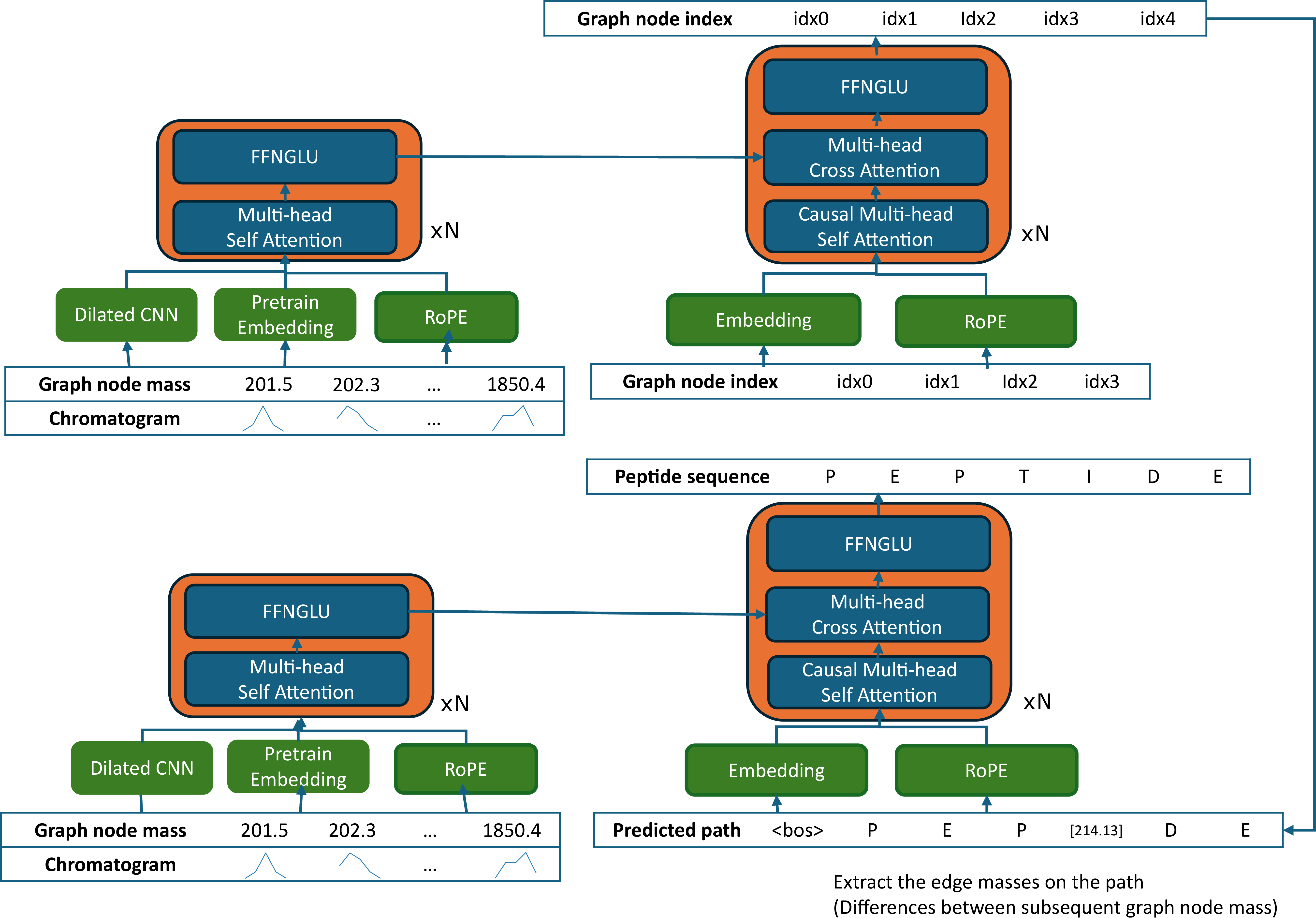}
    \caption{Model structure.}
    \label{fig:model_arch}
  \end{subfigure}
  \hfill
  \begin{subfigure}[b]{0.35\textwidth}
    \includegraphics[width=\textwidth]{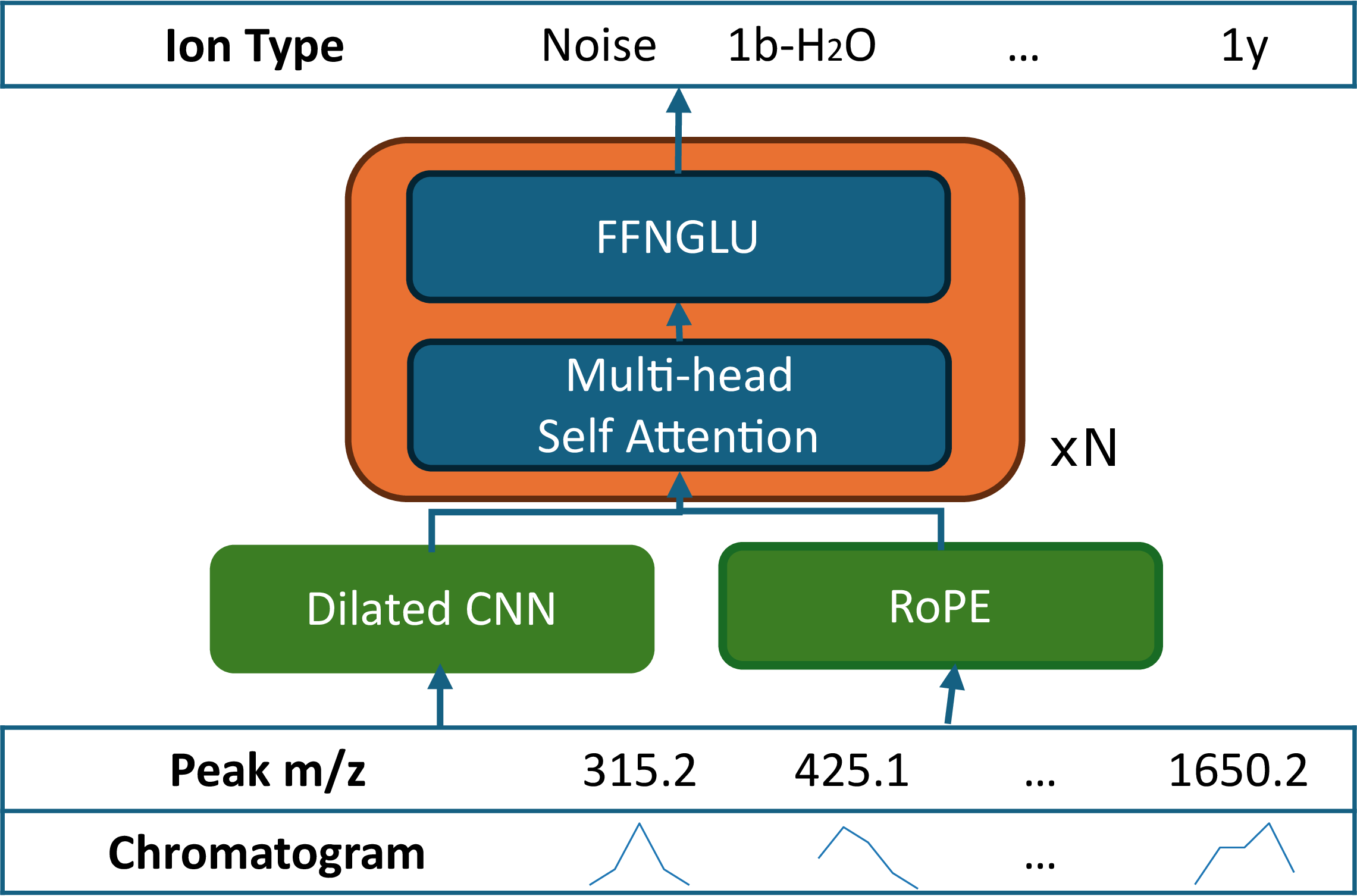}
    \caption{Pretrain model.}
    \label{fig:coelution_pretrain}
  \end{subfigure}
  \hfill
  \vspace{1em}

  \begin{subfigure}[b]{0.76\textwidth}
    \includegraphics[width=\textwidth]{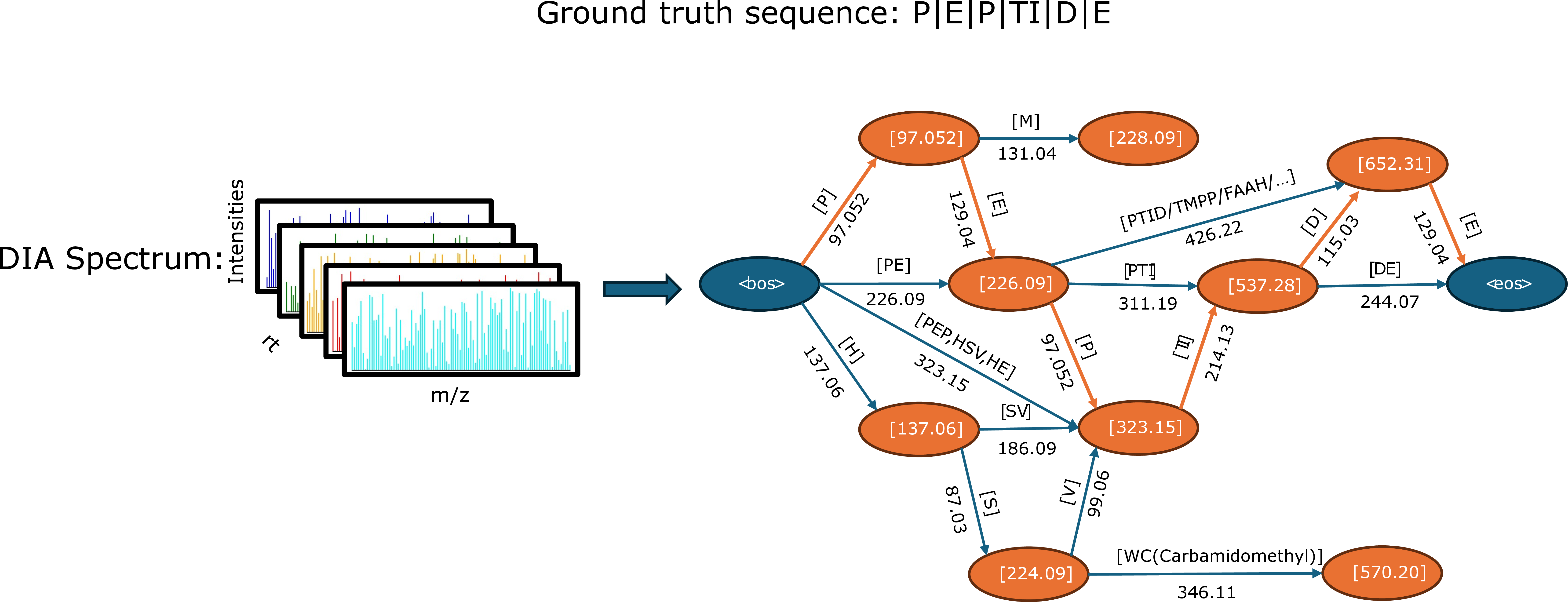}
    \caption{An example of a spectrum graph.}
    \label{fig:spectrum_graph}
  \end{subfigure}

\end{minipage}

\caption{(a) The model structure of Our entire workflow. On the top is the optimal path task, generating a series of node indices, transformed into the optimal path. The mass values in the optimal path are then translated to the corresponding amino acids when a single match is found. On the button is the sequence generation task. It takes the generated optimal path and outputs the amino acid sequence to replace mass tags. (b) The BERT\cite{devlin2019bert}-like pretrain model (c) An example of a spectrum graph, where the bottom value on each edge represents the mass difference between nodes, encoded by RoPE, and the top value indicates the corresponding amino acid sequence. Only a subset of nodes and edges is plotted for clarity, whereas, in a complete spectrum graph, all possible forward connections would be present.}
\label{fig:model_overall}
\end{figure}

We propose DIANovo, a framework specifically designed to address the complexities of DIA data introduced by coelution, with its architecture shown in Figure \ref{fig:model_arch}. Concretely, we first set to tackle the sheer size of highly-multiplexed DIA spectrum. Our encoder encodes the spectrum graph, like GraphNovo\cite{mao2023graphnovo}, as shown in Figure \ref{fig:spectrum_graph}. To reduce memory consumption, we implements an automated edge generation process, which primarily focuses on learning the relative mass differences between graph nodes while the inter-node amino acid information is learned in a task-specific manner. Moreover, we introduce a coelution-aware pretraining step to incorporate information about coeluting peptides alongside the target peptide, shown in Figure \ref{fig:coelution_pretrain}. This involves pretraining a model to predict ion types from coeluting peptides, with the resulting embeddings used as features in subsequent training stages. The coelution information helps the model better differentiate between signal and noise, leading to more accurate target peptide predictions.

Additionally, our study seeks to assess the realistic performance of DIA de novo sequencing, by reporting amino acid and peptide recalls and precisions compared with DIA-NN \cite{demichev2020diann} search outcomes, considering only peptides unique to the test set. Our findings indicate that de novo peptide sequencing using DIA on older-generation instruments like the Q Exactive\cite{michalski2011qexactive} or Fusion \cite{wenger2011fusion} is suboptimal, yielding relatively lower peptide recall. However, the Orbitrap Astral \cite{guzman2024astraldia} significantly outpaces older-generation in acquisition speed, enabling the use of narrow-window data-independent acquisition (nDIA). The Astral narrow-window DIA method \cite{guzman2024astraldia}, which offers more consistent fragmentation patterns and fewer missing fragmentations, even in case of high coelution level, yields better results. Our extensive experiments across various datasets, including older-generation and Astral data, highlight the robustness and effectiveness of our proposed method. We demonstrated that our model consistently outperformed the baselines. These findings underscore the potential of our method for robust and accurate de novo peptide sequencing in the challenging DIA setting.

To better understand these challenges, we developed a theoretical framework that explains how the balance between signal enhancement and noise accumulation in different acquisition methods affects de novo peptide sequencing performance. Specifically, we analyze the signal and noise characteristics of DDA, older-generation DIA, and Astral DIA, and their impact on the efficacy of peptide matching algorithms like XCorr.\cite{eng1994sequest} \cite{noble2012pvalue} Our model reveals that older-generation DIA introduces substantial noise that outweighs the marginal gain in signal peaks—diminishing de novo sequencing performance—while Astral DIA provides a disproportionate increase in signal peaks despite higher noise levels. This increase effectively enhances peptide identification by improving the confidence of peptide-spectrum matches. This theoretical insight underscores the critical importance of optimizing signal-to-noise profiles in mass spectrometry data to improve de novo sequencing outcomes.

Finally, we address a key question: can de novo sequencing in DIA mode detect more peptides than DDA mode? We present a comparison between DDA and DIA data acquired from the same biological sample, demonstrating that with older-generation mass spectrometers, DIA surpasses DDA in peptide detection when using smaller isolation windows. However, as the isolation windows widen, DIA loses this advantage and falls behind DDA. With Astral data, DIA can consistently detect more peptides than DDA because of the narrow DIA method enabled, showcasing the superior performance of next-generation mass spectrometry.

\section{Results}
Our experiments comprehensively evaluated the performance and robustness of our de novo peptide sequencing algorithm across various datasets and conditions. We found that older-generation mass spectrometers exhibit relatively lower peptide recall, while Astral data showed better results due to improved fragment ion coverage. Our model consistently outperformed the baselines DeepNovo-DIA and Transformer-DIA model across multiple datasets, highlighting its superior capability in complex DIA settings. We also include a comparison with a recent state-of-the-art model, Cascadia, demonstrating the superior performance of our approach.

Furthermore, in the comparison of DDA and DIA on the Hela~\cite{ting2017pecan} (older-generation) and PXD046453 \cite{guzman2024astraldia} Hek293T (Astral) dataset, DIA demonstrated advantage in peptide detection when isolation window is small, with our de novo sequencing algorithm identifying additional peptides that DDA missed. These findings provide valuable insights into the proper scenarios for applying DIA de novo sequencing and highlight DIA's capability to extend identification beyond DDA. 

Finally, to explain the observed performance, we propose a theoretical analysis of the relationship between spectrum characteristics and peptide-matching confidence.

\subsection*{De Novo Performance on Older-Generation Data}

\begin{figure}[t!]
\centering
\begin{subfigure}[b]{0.28\textwidth}
   \includegraphics[width=\textwidth]{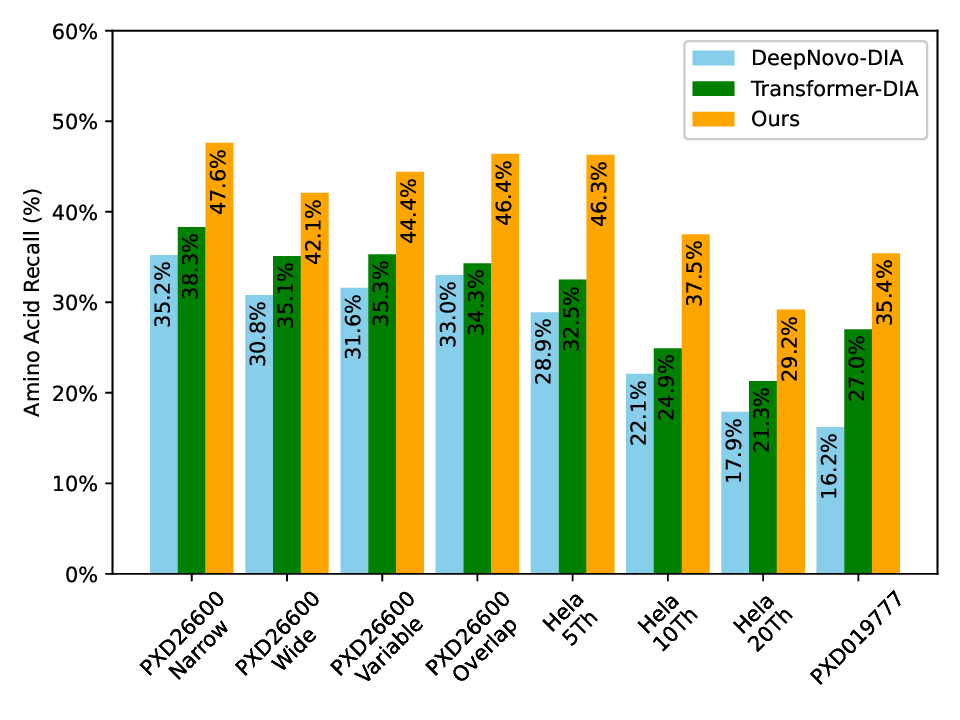}
   \caption{Amino acid recall}
   \label{fig:aa_orbitrap} 
\end{subfigure}
\hfill
\begin{subfigure}[b]{0.28\textwidth}
   \includegraphics[width=\textwidth]{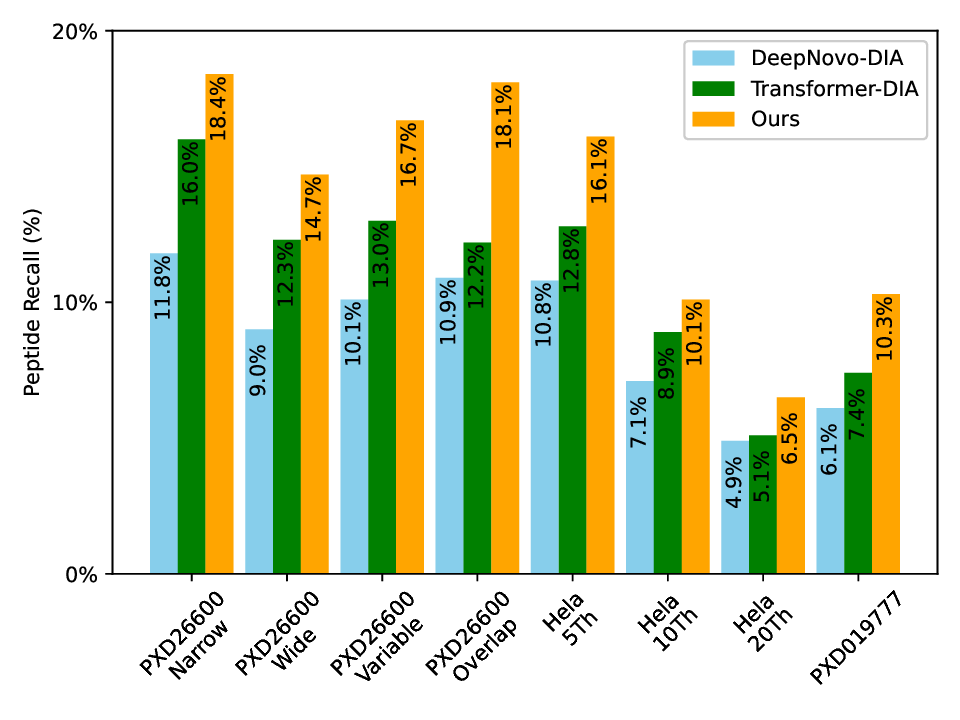}
   \caption{Peptide recall}
   \label{fig:pep_orbitrap}
\end{subfigure}
\hfill
\begin{subfigure}[b]{0.28\textwidth}
   \includegraphics[width=\textwidth]{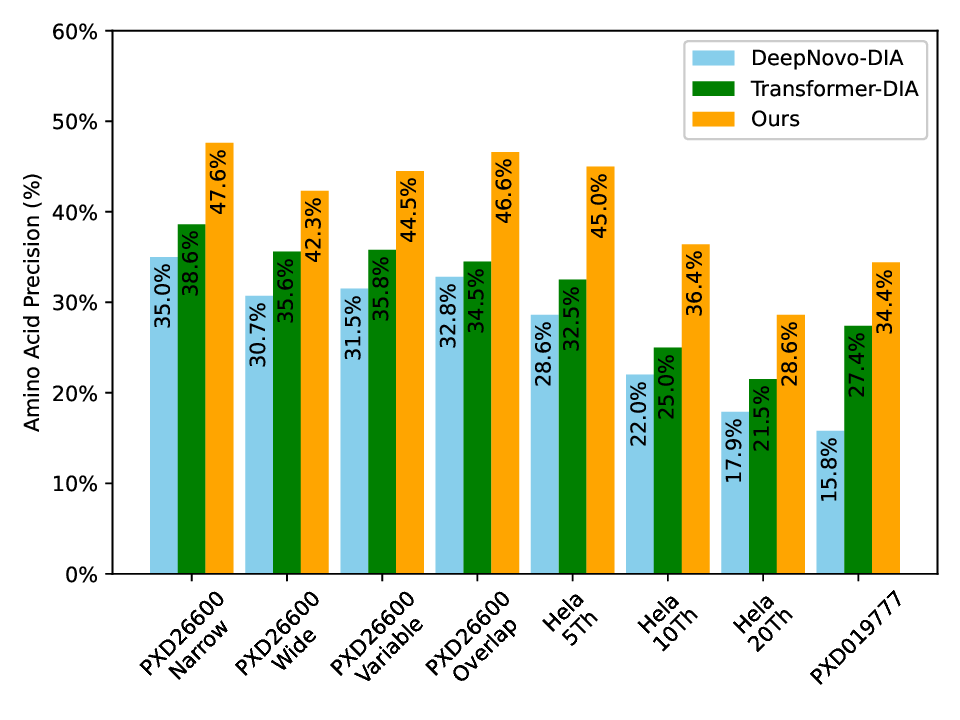}
   \caption{Amino acid precision}
   \label{fig:prec_orbitrap}
\end{subfigure}
\caption{Amino acid recall (a), peptide recall (b), and amino acid precision (c) of our method vs DeepNovo-DIA and Transformer-DIA, training sequences excluded from test set, on various older-generation datasets.}
\label{fig:orbitrap}
\end{figure}

We compared our model's performance with the baselines across several datasets \cite{kalxdorf19777} \cite{gotti26600} \cite{ting2017pecan} in Figure \ref{fig:orbitrap}, where peptide recall is defined as the proportion of peptides whose entire sequences exactly match the database search results, without any mismatches or errors. On most datasets, our model demonstrated significantly better performance than the baselines, however the results are relatively lower across the board compared to DDA levels \cite{tran2017deepnovo} \cite{mao2023graphnovo}, indicating that there is a significant gap between database and de novo identification performance in older-generation mass spectrometers. Our amino acid recall is on average 60\% higher than DeepNovo-DIA, while peptide recall being 53\% higher. Comparing to Transformer DIA, our amino acid recall and peptide recall is 32\% and 24\% higher respectively.

Although our method shows greater ability beyond the baselines, the results indicate that further refinements are needed to achieve satisfactory accuracy in de novo peptide sequencing in older-generation instruments. 

\subsection*{Performance on Orbitrap Astral Data}

\begin{figure}[t!]
\centering
\begin{subfigure}[b]{0.28\textwidth}
   \includegraphics[width=\textwidth]{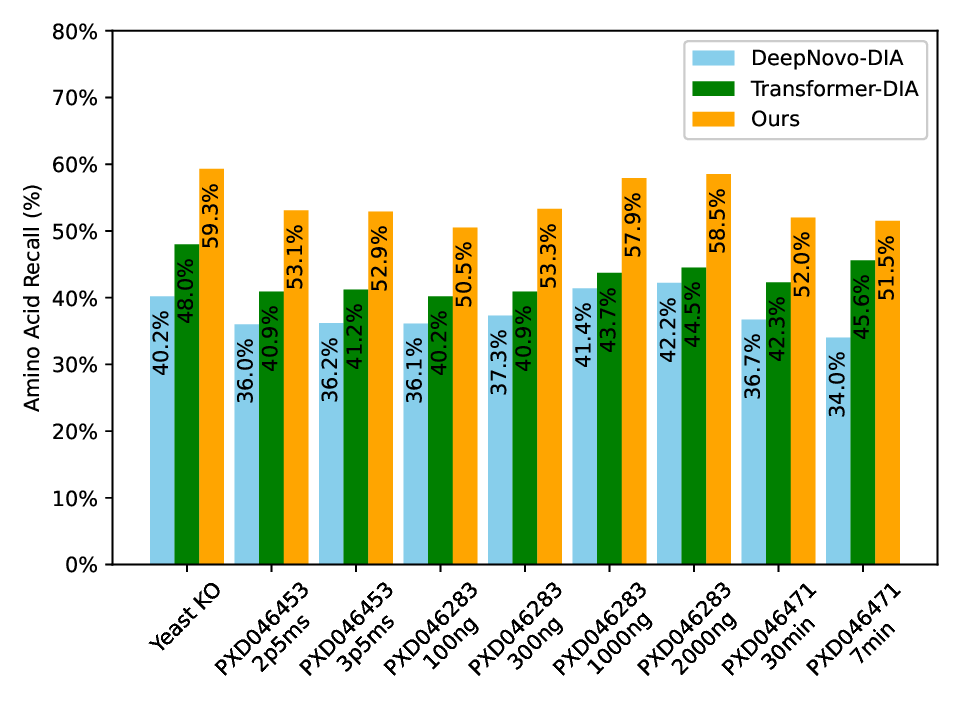}
   \caption{Amino acid recall}
   \label{fig:aa_ast} 
\end{subfigure}
\hfill
\begin{subfigure}[b]{0.28\textwidth}
   \includegraphics[width=\textwidth]{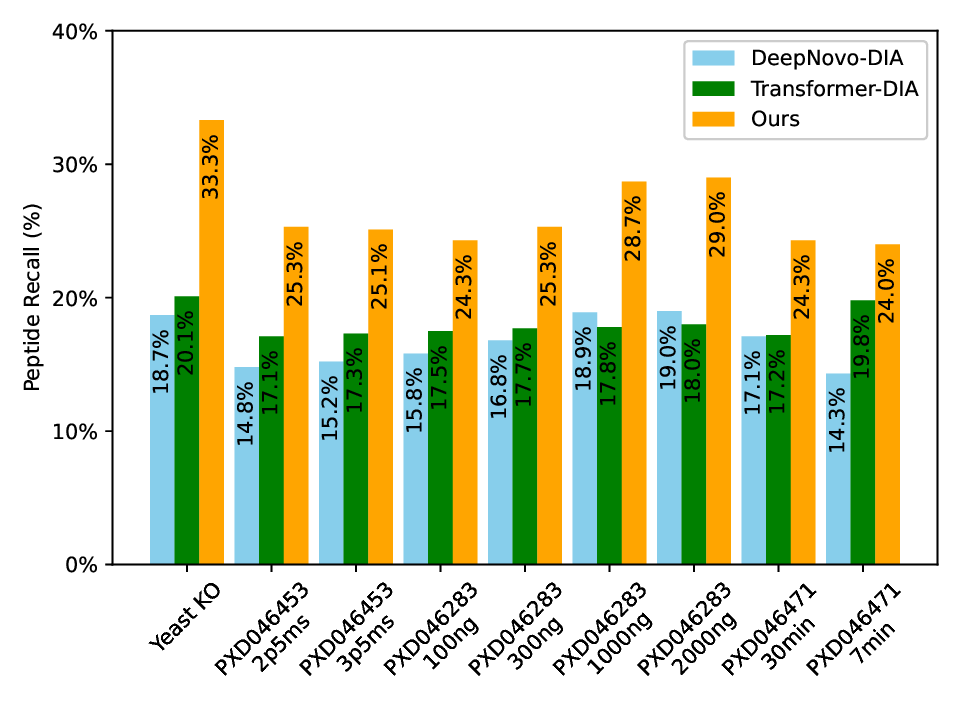}
   \caption{Peptide recall}
   \label{fig:pep_ast}
\end{subfigure}
\hfill
\begin{subfigure}[b]{0.28\textwidth}
   \includegraphics[width=\textwidth]{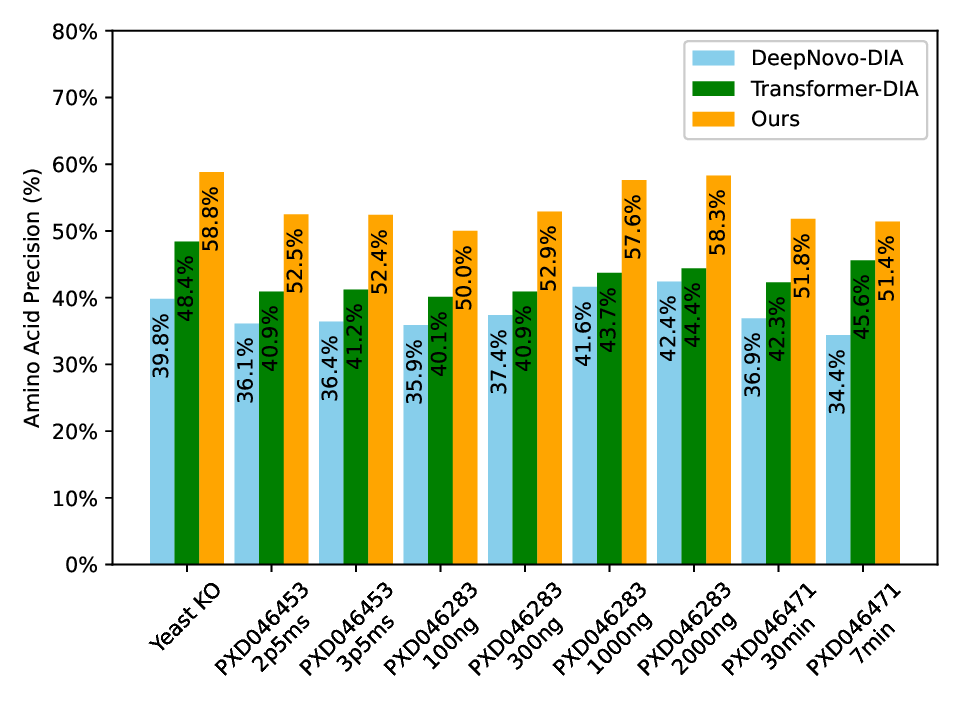}
   \caption{Amino acid precision}
   \label{fig:prec_ast}
\end{subfigure}
\caption{Amino acid recall (a), peptide recall (b), and amino acid precision (c) of our method vs baselines on various Astral datasets.}
\label{fig:astral}
\end{figure}

Our findings indicate that Astral data outperforms older-generation data in de novo peptide sequencing. This superior performance can be attributed to better fragment ion coverage in the Astral data, resulting in fewer missing fragments during the sequencing process. The performance differences are visually represented in the Figure \ref{fig:astral}, comparing our model's performance against the baseline across the different datasets.

The performance of our model on the Astral datasets shows a marked improvement over the baselines. On average we can expect a 43\%/60\% increase of amino acid / peptide recall with our methods compared to DeepNovo-DIA, or a 27\%/47\% improvement compared to Transformer-DIA. Furthermore, Astral data delivers a significant boost to de novo performance, affirming the effectiveness of DIA de novo sequencing in such system.

\subsection*{Comparison of Peptide Detection by DDA and DIA}

\begin{figure}[t!]
\centering
\begin{subfigure}[b]{0.45\textwidth}
   \includegraphics[width=\textwidth]{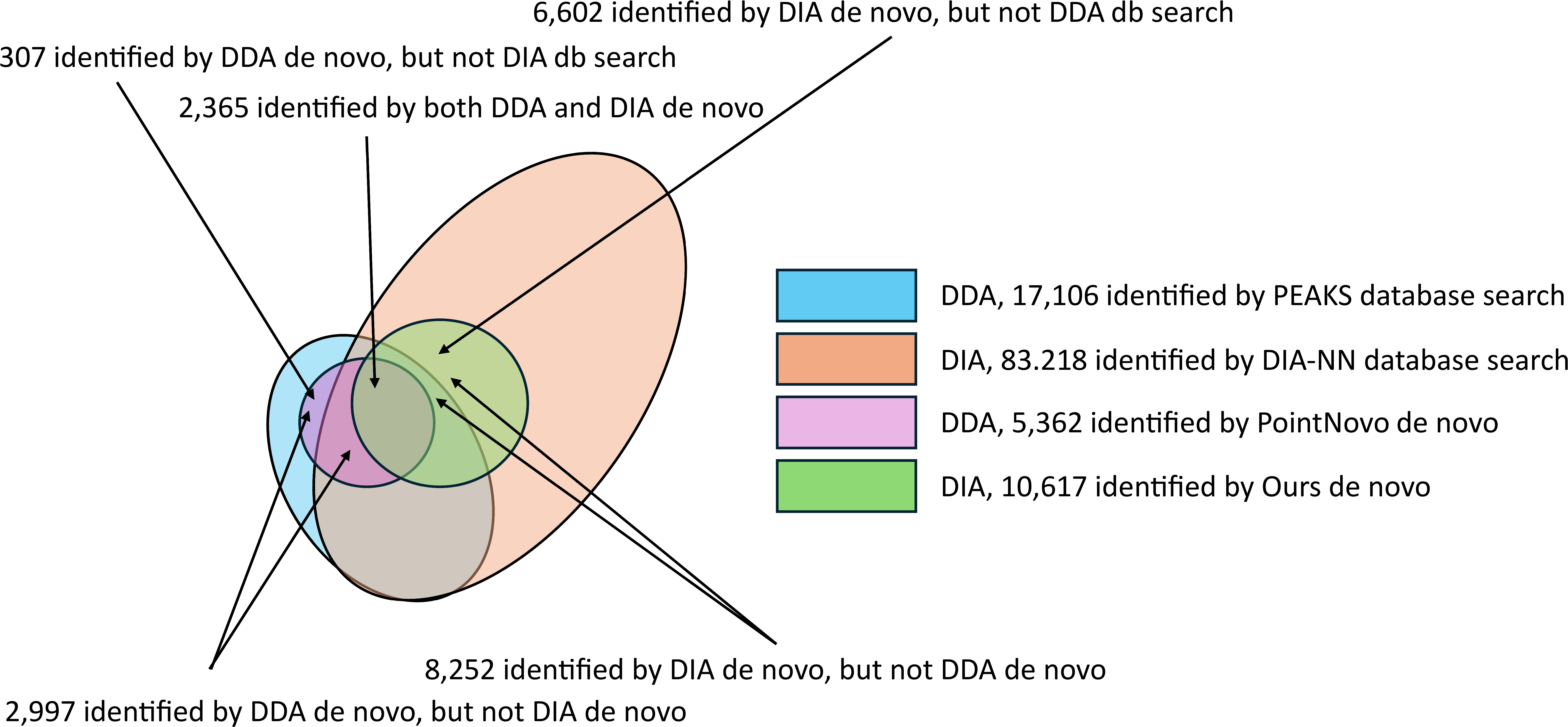}
   \caption{5 Th DIA window}
   \label{fig:5da_hela} 
\end{subfigure}
\hfill
\begin{subfigure}[b]{0.45\textwidth}
   \includegraphics[width=\textwidth]{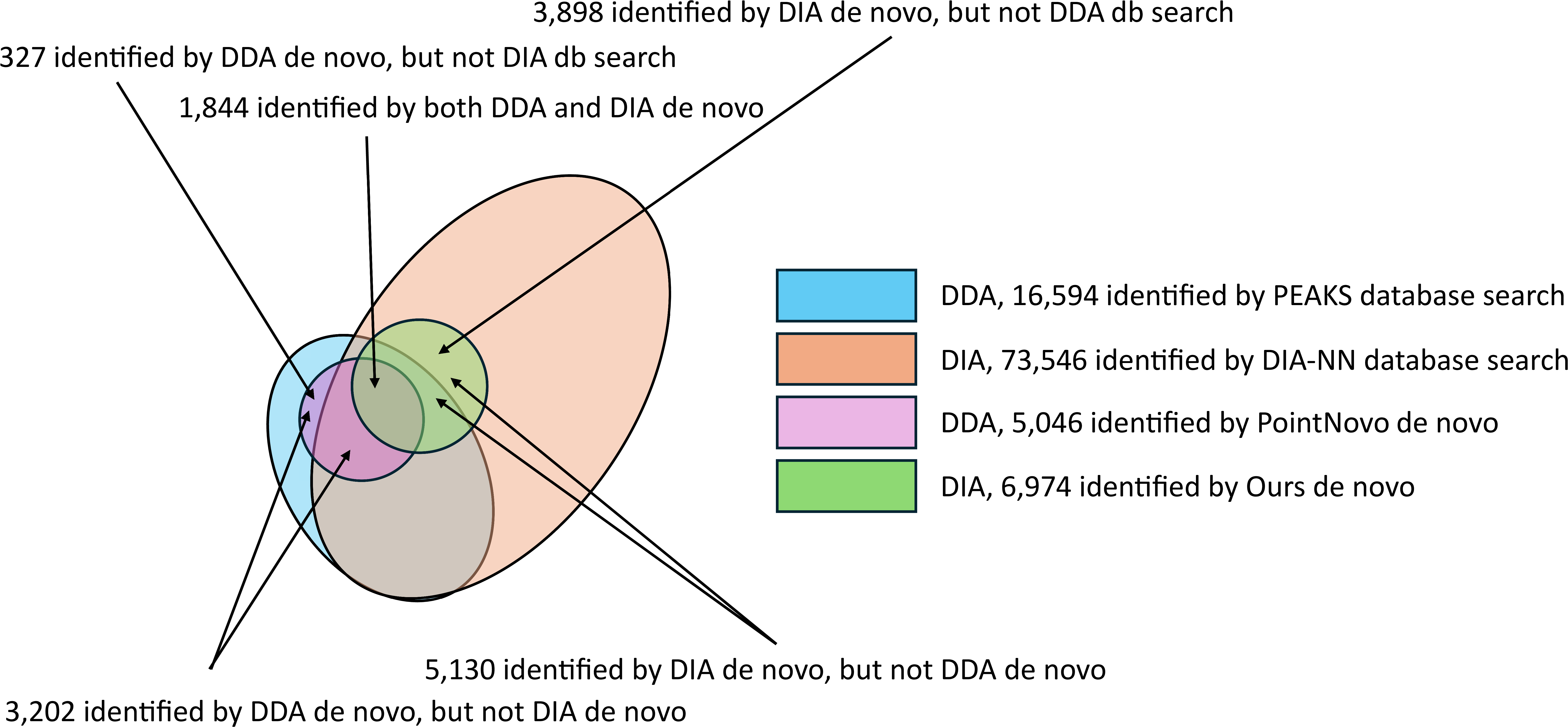}
   \caption{10 Th DIA window}
   \label{fig:10da_hela}
\end{subfigure}

\begin{subfigure}[b]{0.45\textwidth}
   \includegraphics[width=\textwidth]{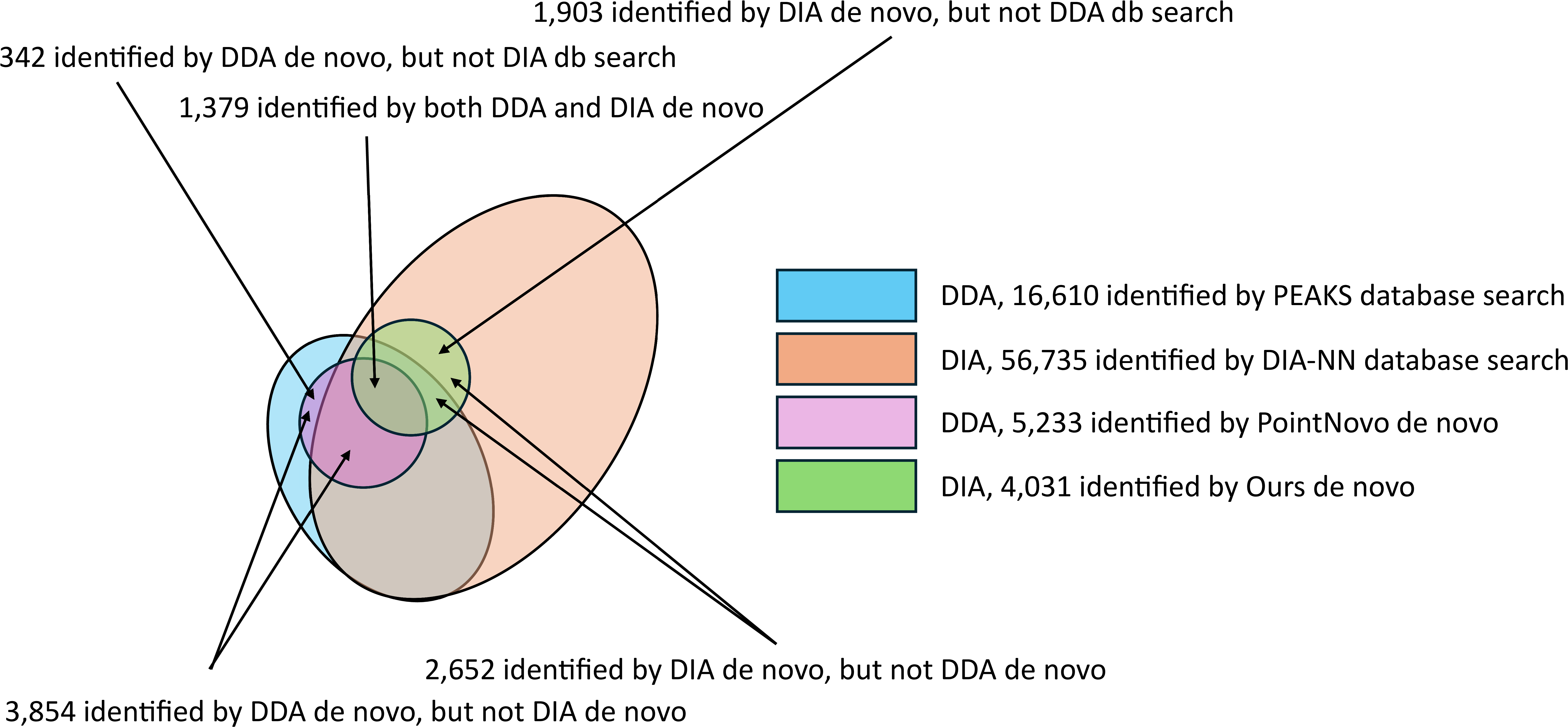}
   \caption{20 Th DIA window}
   \label{fig:20da_hela}
\end{subfigure}

\caption{Venn diagram, comparison of peptide identification under DDA or DIA mode, with Orbitrap Q Exactive (older-generation).}
\label{fig:dda_dia_orbitrap}
\end{figure}

\begin{figure}[t!]
\centering
\begin{subfigure}[b]{0.45\textwidth}
   \includegraphics[width=\textwidth]{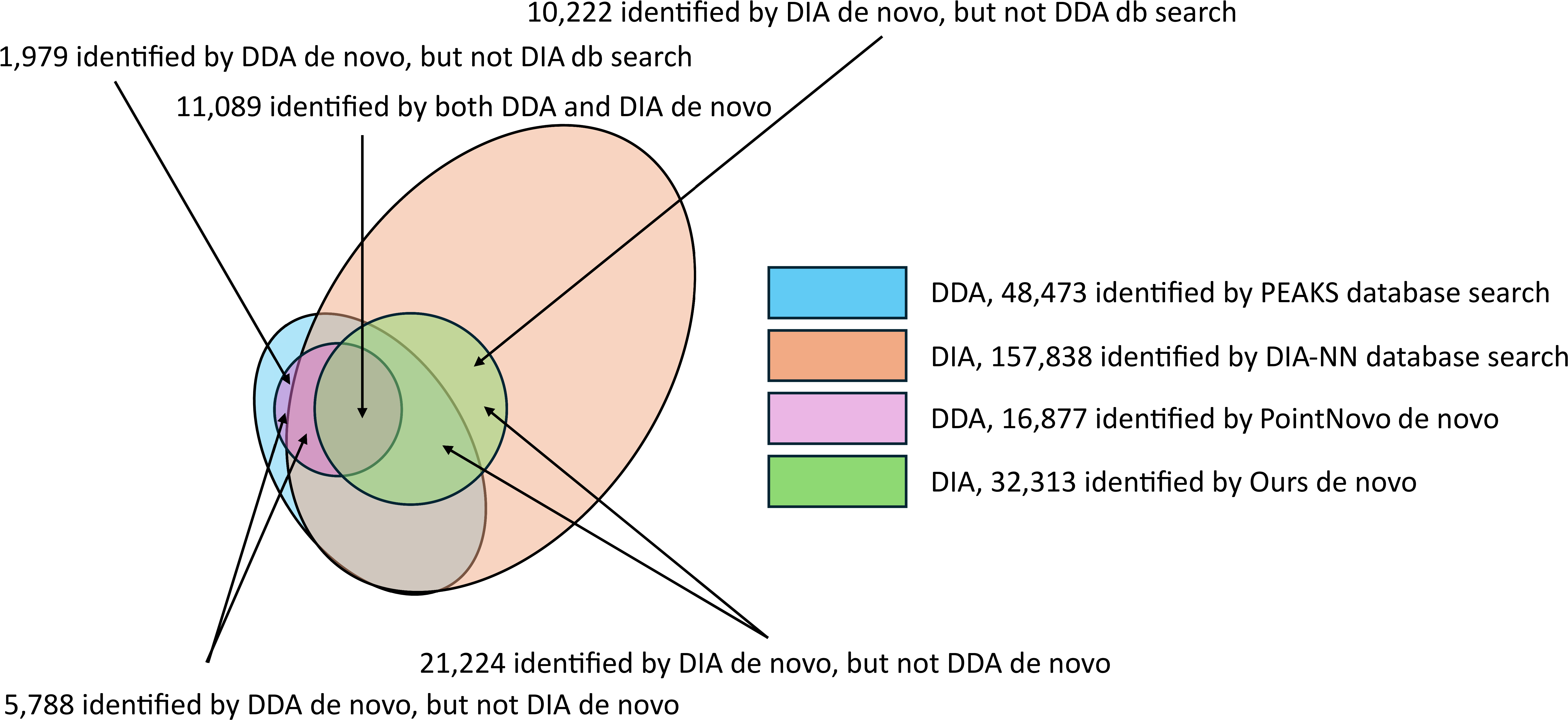}
   \caption{2.5ms DIA Data}
   \label{fig:2p5ms} 
\end{subfigure}
\hfill
\begin{subfigure}[b]{0.45\textwidth}
   \includegraphics[width=\textwidth]{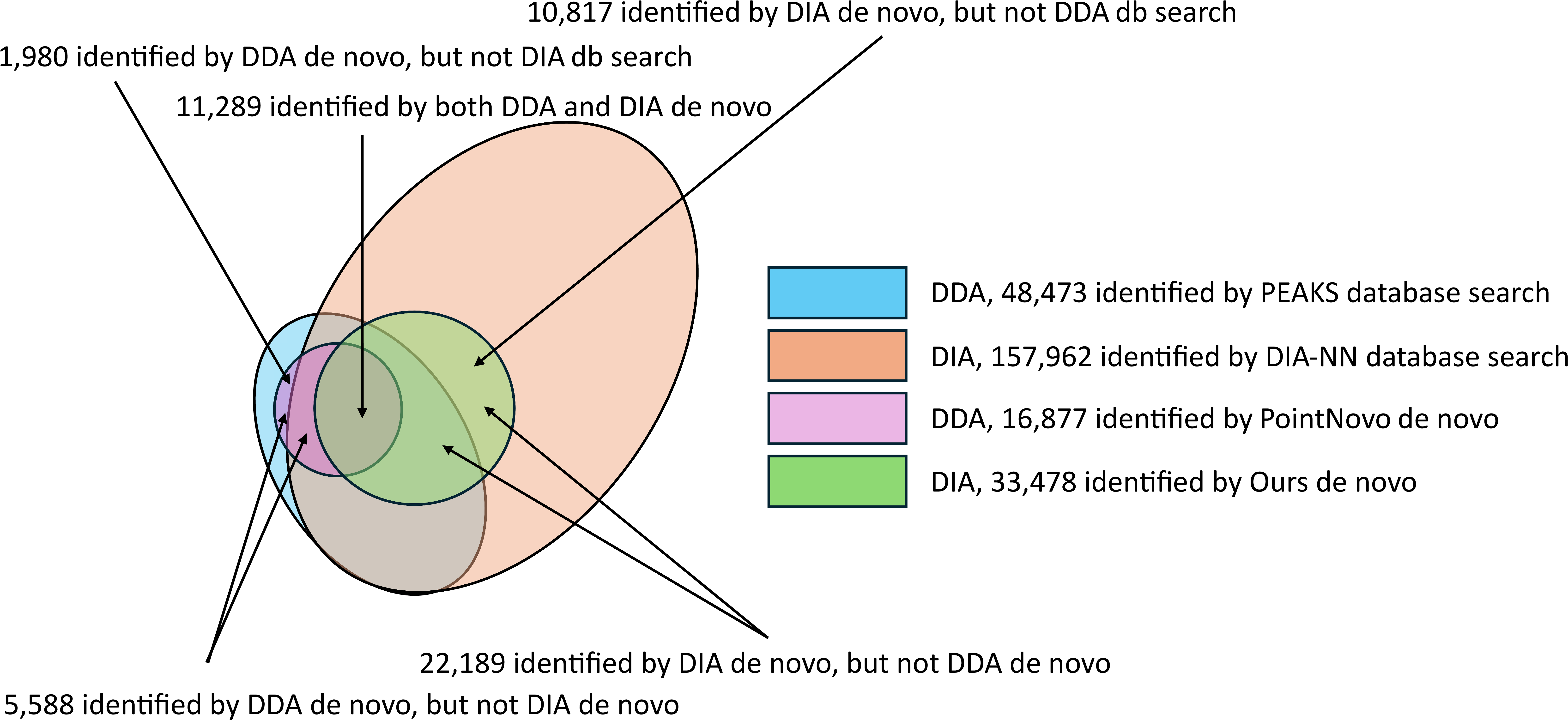}
   \caption{3.5ms DIA Data}
   \label{fig:3p5ms}
\end{subfigure}
\caption{Venn diagram, comparison of peptide identification Under DDA or DIA mode, with Orbitrap Astral.}
\label{fig:dda_dia_astral}
\end{figure}

In this experiment, we conducted a detailed comparison of the number of peptides detected by de novo sequencing using both Data-Dependent Acquisition (DDA) and Data-Independent Acquisition (DIA) modes, utilizing both older-generation mass spectrometers and the newer Orbitrap Astral model. The goal was to understand how these acquisition strategies perform across different experiment settings, particularly in the context of de novo peptide identification.

For older-generation instruments, we adopted the Hela dataset \cite{ting2017pecan}, analyzing peptides identified from the same biological sample in the DDA and DIA modes with varying DIA isolation windows (5 Th, 10 Th, and 20 Th). In the DDA experiments, we performed a database search using the PEAKS\cite{zhang2012peaks}  DB search engine and employed PointNovo~\cite{qiao2021pointnovo} for de novo peptide sequencing. For DIA, we utilized the DIA-NN\cite{demichev2020diann} search engine for database search and applied our de novo sequencing methods.

The results (shown in Figure \ref{fig:dda_dia_orbitrap}) reveal a clear relationship between the isolation window size and the efficacy of peptide detection. When the isolation window is narrow (5 Th), DIA de novo sequencing significantly outperforms DDA, identifying almost twice as many peptides. This suggests that in older-generation mass spectrometers, smaller isolation windows in DIA mode allow for more precise detection, enhancing the depth of peptide identification. However, as the isolation window increases to 10 Th and 20 Th, the performance of DIA begins to diminish in both de novo sequencing and database search, losing its advantage over DDA. At a 20 Th isolation window, while DIA database search still detects more peptides than DDA, de novo sequencing lags behind. The additional peptides identified in database search mode cannot be accurately recovered by de novo sequencing. This decline illustrates that larger isolation windows introduce higher-level of coelution, leading to unwanted noise and overlapping ion signals, reducing the accuracy and efficiency of peptide detection. Therefore, increasing the isolation window in older-generation mass spectrometers appears to be a suboptimal approach for DIA experiments.

In contrast, the performance of the Orbitrap Astral mass spectrometer, using the PXD046453 \cite{guzman2024astraldia} HEK293T dataset, presents a more consistent scenario (shown in Figure \ref{fig:dda_dia_astral}) due to the enabled narrow window mode. This dataset provides both DDA data and DIA data acquired at different cycle times (2.5 ms and 3.5 ms), both using a narrow 2 Th isolation window. The Astral data demonstrates that DIA consistently outperforms DDA with the narrow window mode enabled. The DDA de novo sequencing identifies approximately 17,000 peptides, whereas the DIA de novo sequencing detects nearly 32,000 peptides, showing a considerable increase over DDA. The additional ~15,000 peptides identified by DIA de novo are likely missed by DDA due to biased sampling. This highlights the key advantage of DIA over DDA—its comprehensive and unbiased sampling of peptides, which is further enhanced by the speed and sensitivity of the Orbitrap Astral. In this case, the smaller isolation window and improved performance of Astral’s DIA mode enable the identification of a broader range of peptides, overcoming the limitations observed with DDA.

Overall, the results of both datasets underscore the advantages of using DIA de novo sequencing with narrow isolation windows, particularly when coupled with the advanced capabilities of next-generation instruments like the Orbitrap Astral. This mode not only boosts peptide detection rates but also minimizes the sampling bias inherent in DDA, providing a more comprehensive view of the proteome.

\subsection*{Comparison with Cascadia \cite{sanders2024cascadia}}

\begin{figure}[t!] 
\centering \includegraphics[width=0.45\textwidth]{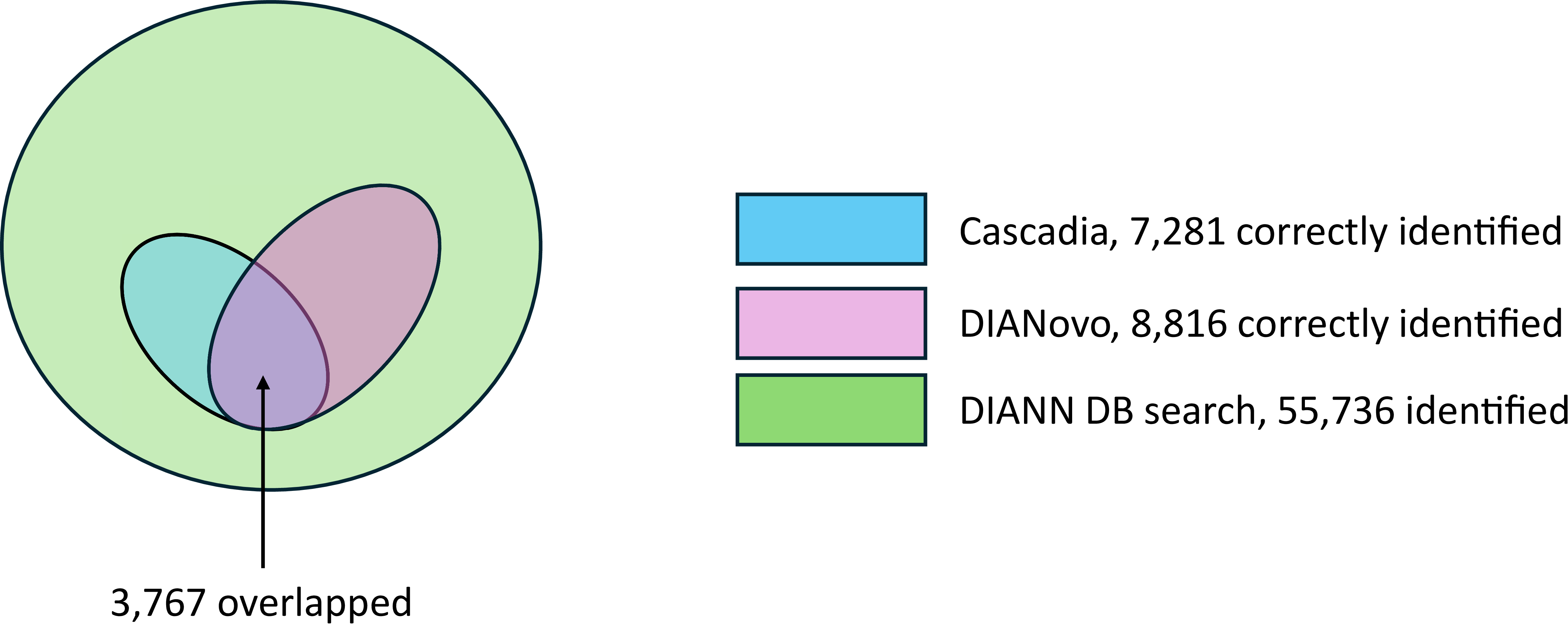} 
\caption{Venn diagram for Cascadia comparison.} 
\label{fig:cascadia} 
\end{figure}

To evaluate the performance of our approach in a practical setting, we conduct a comparative analysis against Cascadia, a recently introduced state-of-the-art model. For this purpose, we selected a representative raw file from the PXD046386 Yeast Knockout (KO) dataset. This dataset provides a relevant benchmark for assessing model effectiveness in proteomics data analysis. Both our method and Cascadia were applied to this identical data source under equivalent preprocessing and parameter settings to ensure a fair comparison. The resulting performance differences are summarized and visualized in Figure \ref{fig:cascadia}.

Our model demonstrates a 21\% increase in peptide identifications compared to Cascadia. Notably, the overlap in identified peptides between the two models is relatively limited, indicating that each model captures distinct subsets of the data. This suggests that our approach exhibits different identification characteristics, potentially offering complementary insights to those provided by Cascadia.

\subsection*{Theoretical Analysis of De Novo Peptide Sequencing Performance}
We present a theoretical framework to explain the performance variations observed in de novo peptide sequencing across different mass spectrometry acquisition methods: Data-Dependent Acquisition (DDA), older-generation Data-Independent Acquisition (DIA), and Astral DIA. Our analysis focuses on how the balance between signal and noise in the generated spectra affects the efficacy of a generic peptide matching algorithm XCorr, illustrating the effect of signal and noise profile on peptide spectrum match quality, resulting in different de novo sequencing or database search performance.

\begin{figure}[t!]
\centering
\begin{subfigure}[b]{0.35\textwidth}
\includegraphics[width=\textwidth]{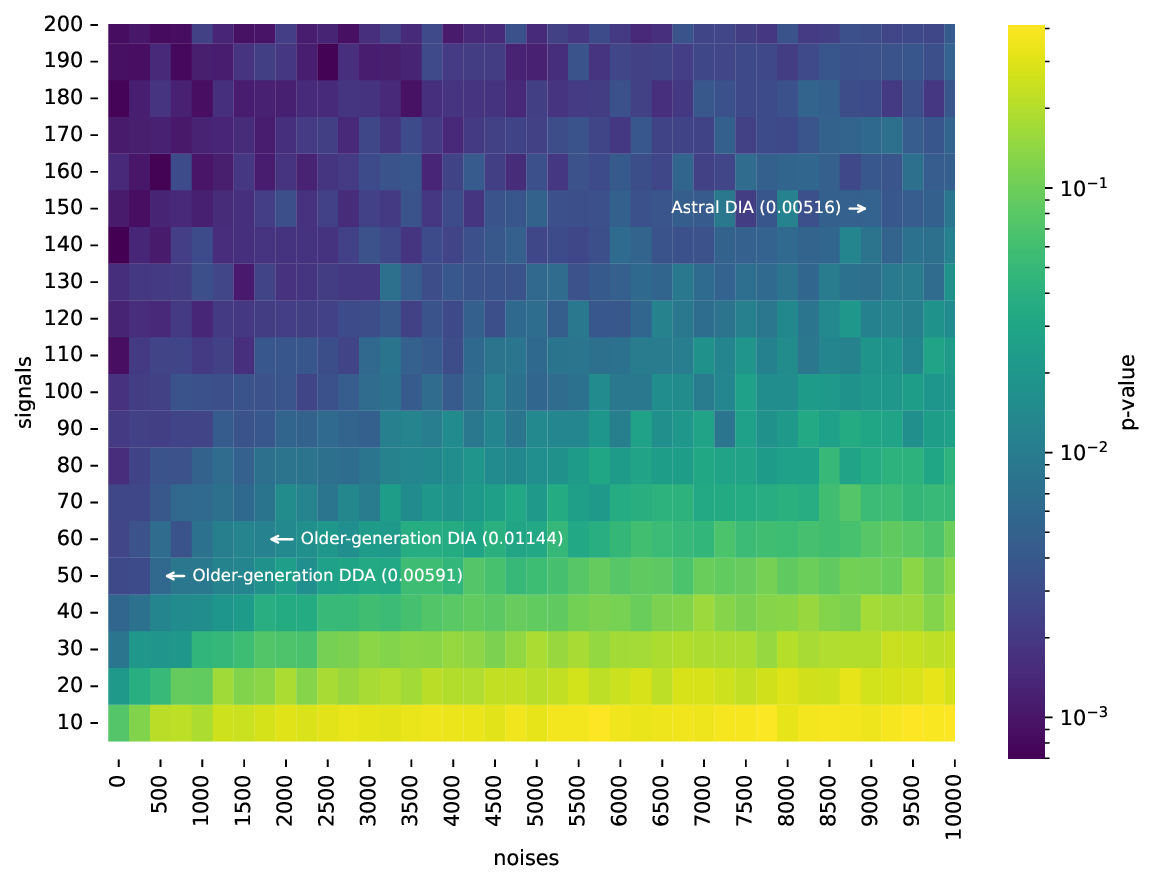} 
\caption{}
\label{fig:heatmap} 
\end{subfigure}
\begin{subfigure}[b]{0.4\textwidth}
\includegraphics[width=\textwidth]{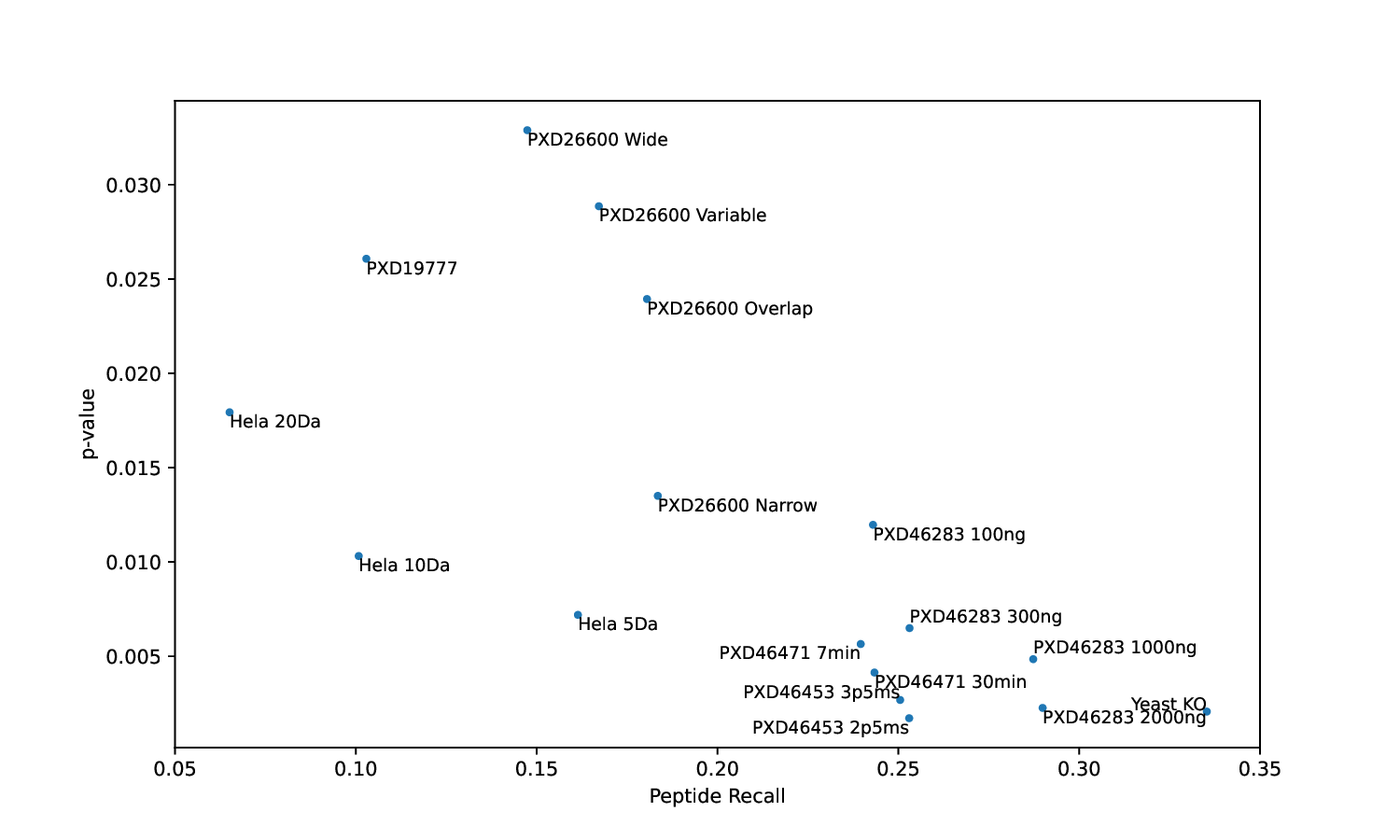} 
\caption{} 
\label{fig:pvalue_vs_peprecall} 
\end{subfigure}
\label{fig:theory_graph}
\caption{(a) Simulated p-values for different signal and noise values. In the figure, older-generation DDA points to 50 signal peaks and 500 noise peaks with p-value of 0.00591, older-generation DIA points to 60 signal peaks and 1,750 noise peaks with p-value 0.01144, while Astral-DIA points to 90 signal peaks and 9,000 noise peaks with p-value 0.00516. (b) Simulated p-values vs peptide recall for test datasets, with a Pearson correlation coefficient of -0.68.}
\end{figure}

\subsubsection*{Signal and Noise Characteristics in Different Acquisition Methods}

We define signal peaks as the fragment ions originating from the target peptide. All other peaks are considered noise, which typically includes fragment ions from coeluting peptides, immonium ions, and instrumental noise. In DDA spectra, we typically observe approximately 50 signal peaks corresponding to fragment ions of the peptide of interest, along with about 500 noise peaks. Transitioning from DDA to older-generation DIA results in a slight increase in signal peaks to around 60 but introduces a substantial increase in noise peaks to approximately 1,750, due to spectra being highly-multiplexed. We argue that this significant escalation in noise outweighs the marginal gain in signal peaks, leading to diminished de novo sequencing performance in older-generation DIA compared to DDA.

When moving from older-generation DIA to Astral DIA, one might expect the scenario to revert to that of DDA due to the employment of narrower isolation windows. Contrary to this expectation, Astral DIA produces a considerable increase in signal peaks to approximately 150 and an even larger surge in noise peaks to around 9,000, while demonstrating high level of coelution in spite of the narrow isolation window. Additionally, Astral DIA data exhibits lower noise peak intensity relative to signal peaks compared to older-generation equipment. Despite the higher noise levels, Astral DIA demonstrates significantly better de novo sequencing performance than older-generation DIA.

\subsubsection*{Theoretical Model Explaining the Observed Performance}

To elucidate the observed performance, we propose a theoretical model based on the difficulty of matching peptides against noise using the widely adopted peptide matching algorithm, XCorr. Our central argument is that a higher XCorr value for a de novo peptide indicates a more straightforward and confident match of the peptide sequence in both database searches and de novo sequencing algorithms. 

In database searches, the XCorr score \cite{eng1994sequest} quantifies the similarity between the experimental spectrum and theoretical spectra generated from candidate peptides in the database. A higher XCorr score reflects a better match, leading to increased confidence in peptide identification. This is because the probability of a high-scoring match occurring by chance decreases exponentially with increasing score, enhancing the specificity of the identification. For de novo sequencing, the advantage of a higher XCorr score stems from the intrinsic similarities between database search algorithms and de novo methods. De novo algorithms are heuristic approaches designed to reconstruct peptide sequences directly from spectra without relying on a predefined database. Although de novo methods may employ advanced scoring functions tailored for sequence reconstruction, the XCorr score serves as a valuable lower-bound approximation of their performance. A higher XCorr score implies that the spectrum contains clear and informative fragment ion peaks that can be effectively utilized by de novo algorithms to deduce the peptide sequence.

\subsubsection*{Simulation of Experimental Scenarios}

To validate our theory that signal-noise profile plays a crucial rule in de novo performance, we conducted simulations by performing a grid search through signal peak and noise peak numbers by generating synthetic spectra. Utilizing the approach proposed by Noble et al. (2012)\cite{noble2012pvalue}, we computed the p-value for a single peptide-spectrum match based on dynamic programming and XCorr scoring. This method allows for the estimation of the statistical significance of the observed XCorr scores, providing insight into the confidence of peptide identifications under varying signal-to-noise conditions. 

Our simulation results are presented as a heatmap in Figure \ref{fig:heatmap}, illustrating the impact of the signal-to-noise profile on matching quality. The three experimental scenarios DDA, older generation DIA, and Astral DIA are labeled within the figure for reference. The simulations reveal that the computed p-values align with our theoretical expectations. Specifically, both older-generation DDA and Astral DIA exhibit similar p-values, which are significantly lower than those observed for older-generation DIA. This finding suggests that peptide-spectrum matching is intrinsically more favorable in the contexts of Astral DIA and DDA compared to older-generation DIA. The improved performance of Astral DIA compared to older-generation methods demonstrates that, despite its higher noise levels, it also produces significantly more signal peaks. In this scenario, the increased number of signal peaks effectively compensates for the additional noise. This suggests that the absolute signal peak count plays a more critical role than the signal-to-noise ratio, as Astral DIA yields a lower signal-to-noise ratio than older-generation methods yet still achieves superior results.

\subsubsection*{Relationship Between p-value and Peptide Recall}

In theory, the p-value should exhibit an inverse correlation with peptide recall, as discussed in detail in Supporting Section \ref{sec:supporting_theory}. This theoretical relationship is supported by our simulation results using characteristics of test datasets, as shown in Figure \ref{fig:pvalue_vs_peprecall}. We created synthetic spectra that replicate the signal and noise profiles of each test dataset, accounting for peptide length and noise intensity, thereby simulating real experimental conditions. The figure clearly demonstrates an inverse relationship between peptide recall and p-value, indicating that lower p-values are associated with higher recall rates. Furthermore, the results reveal a clear separation between older-generation DIA and Astral DIA data. Older-generation DIA data predominantly occupies the upper-left quadrant, characterized by lower peptide recall and higher p-values. In contrast, Astral DIA data is concentrated in the lower-right quadrant, reflecting its higher peptide recall and lower p-values. This distinction highlights the superior performance of Astral DIA, which can be attributed to its improved signal-to-noise profile and lower noise intensity, resulting in enhanced data quality and spectral characteristics.

This finding highlights the utility of the simulated p-value as a reliable indicator of de novo sequencing performance. By effectively capturing the underlying relationship between statistical significance and identification accuracy, the p-value provides valuable insights into the quality of peptide identifications. These results further validate the robustness of our simulations and their alignment with real experimental conditions, reinforcing the practical relevance of this theoretical framework.

\section{Discussion}
\subsection*{Proposed DIANovo Algorithm}

This study introduces a robust and highly accurate method for de novo peptide sequencing within the DIA setting, offering substantial improvements over existing approaches. With DIA-oriented design, our model is adept at handling the complex, highly-multiplexed and noisy spectral data inherent to DIA experiments. 

Our comprehensive experiments demonstrate the superior performance of the proposed method, particularly when applied to data from the Orbitrap Astral mass spectrometer. The Astral's enhanced fragment ion coverage and consistent fragmentation patterns provide a solid foundation for our de novo sequencing approach, leading to a marked increase in peptide identification accuracy, even in case of high coelution level. In contrast, older-generation mass spectrometers, such as those used in older DDA and DIA experiments, reveal the limitations of existing de novo sequencing methodologies, especially when larger isolation windows are used. Our DDA vs. DIA comparison study highlights these challenges, showing that while DIA outperforms DDA with narrower isolation windows, it begins to lose this advantage as the window size increases. This issue is particularly evident with older-generation instruments, where DIA performance in both database search and de novo sequencing diminishes at larger window sizes. However, the Orbitrap Astral consistently demonstrates DIA's superiority over DDA, facilitated by the narrow isolation windows enabled by this instrument, underscoring the pivotal role that modern mass spectrometers play in advancing peptide sequencing capabilities.

\subsection*{Theoretical Analysis}
During our theoretical analysis of de novo performance, we show that the balance between signal enhancement and noise accumulation is critical. This outcome underscores the importance of optimizing acquisition parameters to maximize signal quality while managing noise levels. We need to note that de novo peptide sequencing is affected more by the deminishing match quality compared to database search, explaining the wider gap between de novo and database search in wide window settings.

Our findings suggest that acquisition methods like Astral DIA, which can substantially increase signal peaks, despite its higher noise and lower signal-to-noise ratio, are more conducive to de novo peptide sequencing. This has important implications for the development of mass spectrometry techniques and the selection of acquisition parameters in proteomics research. Conversely, older-generation DIA does not offer the same advantage because the modest gain in signal peaks is insufficient to counterbalance the significant escalation in noise peaks, despite their lower signal-to-noise ratio. This results in lower confidence in peptide identifications, as reflected by higher p-values. We can also conclude that identification performance depends on more than just the signal-to-noise ratio, with the absolute strength of signal peaks playing a more decisive role.

\subsection*{Potential Applications of Signal- and Noise-Peak–Based Performance Prediction }

We can estimate per-spectrum signal and noise peak numbers without identification by combining an MS1-based purity cue with complementary b/y-ion evidence. Briefly, we measure how much of the local MS1 signal in a DIA frame comes from the dominant precursor (its purity) and use a small, once-calibrated model (driven by precursor intensity, charge, collision energy, resolution, and injection time/AGC) to predict how many fragment peaks that precursor would produce if isolated; scaling this “fragment budget” by purity yields a physics-based signal estimate. In parallel, we scan the MS2 spectrum for complementary b/y pairs whose masses sum to the precursor’s neutral mass (within tolerance) and that co-elute with the precursor in retention time; the union of peaks in validated pairs gives a second, sequence-agnostic signal estimate. Fusing the two gives the signal-peak count; subtracting from the total peaks yields noise.

Compared to performing a small-scale search, this is real-time and engine-independent, provides quality metrics for all spectra (including those that remain unidentified), avoids circularity with scoring/FDR, and scales to rapid “what-if” evaluations of acquisition settings.

This approach leads to many potential applications. For instance, we can perform real-time/adaptive acquisition: sub-second decisions to extend/shorten injection time, adjust collision energy, skip frames, or schedule narrow-window revisits based on the prediction of spectrum quality we receive. Pre-search compute triage could also be achieved by binning spectra by predicted difficulty and route expensive open-mod/de novo only to spectra likely to succeed, improving IDs per CPU/GPU hour.

\section{Methods}
\label{sec:methods}
\subsection*{DIANovo De Novo Sequencing Model}
Our model architecture is based on an encoder–decoder Transformer, augmented with a BERT-style pretraining module. We operate on a spectrum graph representation, in which each spectrum is converted into an aligned format by grouping peaks across adjacent spectra with closely matching $m/z$ values. This representation naturally encodes chromatographic information along the retention time (RT) dimension, which we find to be highly beneficial for improving de novo sequencing performance. During pretraining, the model takes the original spectrum as input and predicts the ion types for all coeluting peptides, rather than focusing solely on the target peptide, as illustrated in Figure~\ref{fig:coelution_pretrain}. This strategy could enable the model to better capture the multiplexed nature of DIA spectra. For de novo sequencing, we employ a two-stage decoding framework (Figure~\ref{fig:model_arch}). In the first stage, the model predicts the optimal path through the spectrum graph, and in the second stage, it refines this path by filling in mass tags to generate the final peptide sequence. This approach mitigates the sequence memorization problem often observed in de novo sequencing algorithms, where models overly rely on peptides frequently seen during training. Additionally, we incorporate rotary positional embeddings (RoPE) to encode the mass differences between spectrum graph nodes, which we find provides superior performance compared to the use of absolute positional embeddings alone. Further details on the model design are provided in Supporting Section A. 

\subsection*{Theoretical Analysis}
To better understand the statistical behavior of de novo peptide sequencing, we performed controlled simulations to generate synthetic spectra that mimic the signal and noise characteristics of different acquisition methods and test datasets. Signal peaks were sampled from theoretical spectra to represent peptide evidence, while noise peaks were added at levels representative of typical coelution and background noise for each acquisition method. For each simulated spectrum, we computed XCorr scores by cross-correlating the simulated spectra with the corresponding theoretical spectra, followed by significance estimation using a dynamic programming approach for p-value calculation~\cite{noble2012pvalue}. Our analysis further examined the impact of multiple hypothesis testing in de novo sequencing, where the dramatically larger search space compared to database searches necessitates stricter statistical corrections, such as the Šidák adjustment~\cite{sidak1967rectangular}. This adjustment increases the stringency of significance thresholds, which in turn reduces peptide recall in de novo sequencing relative to database-driven methods. This theoretical derivation further illustrates that database searches are less impacted by the elevated p-values arising from the degraded signal-to-noise profiles observed when transitioning from DDA to older-generation DIA. This resilience helps database search methods maintain reasonable performance under noisier conditions, whereas de novo sequencing, with its vastly larger hypothesis space, experiences a more pronounced drop in performance.

A full description of the simulation setup, parameter choices, and theoretical derivations supporting this analysis are provided in the Supporting Materials Section B. 

\section{Conclusion}
In this paper, we present a novel and highly effective method for de novo peptide sequencing in the DIA setting, offering significant improvements over traditional approaches. A key focus of our study is the comparison between DDA and DIA, where we demonstrate that while DIA outperforms DDA with narrow isolation windows, its advantage diminishes with wider windows in older-generation instruments. With Orbitrap Astral,, due to enabled narrow-window mode, DIA consistently surpasses DDA, highlighting the importance of next-generation mass spectrometers in improving peptide identification. To explain this phenomenon, we propose a simulation based theory, and highlight what aspects lead to improved identification performance.

Our method, specifically designed to tackle the highly mutliplexed DIA data, along with leveraging advanced instrumentation, provides robust performance in challenging proteomic environments. These findings emphasize the value of using DIA with modern technologies for comprehensive peptide detection, paving the way for more reliable and effective de novo sequencing methods in proteomics.

\bibliography{reference}

\section*{Acknowledgements}

The authors received support from Canadian Network for Research and Innovation in Machining Technology, Natural Sciences and Engineering Research Council of Canada (NSERC Canadian Network for Research and Innovation in Machining Technology), grant number OGP0046506 [Li].


\section*{Code Availability}

The code for this project is available via \url{https://github.com/hearthewind/dianovo}.


\begin{appendices}
\section{The DIANovo Model}
\label{sec:supporting:model}
\subsection*{Feature Extraction}
\label{sec:spectrum_encoder}
In a tandem mass spectrometry experiment, peptide precursors are first analyzed in their intact form, generating first-stage scans (MS1 scans). The peptides are then fragmented, and the resulting fragments are analyzed again, producing second-stage scans (MS2 scans).

The feature construction for the encoder is designed to capture the chromatogram characteristics in DIA, in order to better distinguish signal peaks from coeluting fragment ions and noise. For each PSM (peptide-spectrum match), we collect five neighboring MS2 spectra along the retention time (RT) dimension, centered around the RT peak. The neighboring spectra provide a chromatogram for each observed peak in the spectrum, and the relationship among these chromatograms improves the robustness of peptide sequence predictions.~\cite{demichev2020diann}. During preprocessing, peaks across neighboring spectra with similar $m/z$ values are grouped together, enabling the construction of chromatograms for each m/z value. This is achieved by binning each spectrum into fixed-size 0.01 Th intervals, and once the chromatograms are generated through binning, the spectra are reconstructed as a list of (m/z, chromatogram) pairs for neural network processing. The binning approach offers greater computational efficiency than methods such as KD-trees, making it particularly well-suited for the characteristics of our DIA data.

For MS1 spectra, alignment with corresponding MS2 spectra is achieved by interpolating MS1 intensity values based on their RT values to match the MS2 spectra, ensuring that the features derived from MS1 and MS2 spectra are directly comparable.

In addition to the primary spectral data, we also included eight supplementary features to enrich the feature set used for encoding, adapted from GraphNovo\cite{mao2023graphnovo} and Novor\cite{ma2015novor}. These features are derived from each spectrum within the $\pm50$ Th neighboring window and are designed to capture various statistical and intensity-based characteristics of the spectral peaks.

Post feature extraction, we compile the data into two structured tensors for each m/z value in combined spectrum:
\begin{itemize}
\item A $[5,1]$ tensor representing the chromatogram data across the five neighboring MS2 spectra. 
\item A $[5,8]$ tensor that encapsulates the eight computed features across the same neighboring spectra. 
\end{itemize}

Finally, the spectrum graph is constructed by transforming every peak in the original spectrum from m/z to N-terminal residual masses. Each peak in the original spectrum is converted into 6 peaks in spectrum graph, taking into account 6 types of ions, including $a^+$, $a^{2+}$, $b^+$, $b^{2+}$, $y^+$, and $y^{2+}$. We operate on the N-terminal residual masses, denoted as the graph node mass, following Equation \ref{eq:graphnode_mass}, during which C-terminal ions are converted to their N-terminal residual masses by subtracting their C-terminal residual mass from the precursor mass. Note that although we did not specifically include other common ion types like $b-H_2O$ or $y-{NH_3}$, their peaks are converted into graph node as well. They do not belong to the target sequence, but their graph node can still contribute to our de novo objective by self attention. We provide an example of spectrum graph in Figure \ref{fig:spectrum_graph}.

\begin{equation}
m_{\text{nterm}} =
\begin{cases}
\left(m/z - m_{\text{proton}}\right) \times c - \sigma,  \\ \hfill \text{if ion} \in \{a, b, c \text{-ions}\} \\
m_{\text{precursor}} - \left(m/z - m_{\text{proton}}\right) \times c + \sigma,  \\  \hfill \text{if ion} \in \{x, y, z \text{-ions}\}
\end{cases}
\label{eq:graphnode_mass}
\end{equation}

\begin{flushleft}
where offset $\sigma$ depends on the ion type, computed as Supporting Table \ref{table:ion_offset}; $c$ is the charge of ion; $m_{proton}$ is the mass of proton and $m_{precursor}$ is the precursor mass~\cite{mao2023graphnovo}. 
\end{flushleft}

\subsection*{Time-Series Encoder and Spectrum Encoder} 
In our approach, we employ a dilated convolutional neural network (CNN) to encode the time-series data derived from the chromatogram and the additional spectral features, encoding time-dependent information in chromatograms while keeping the number of parameters and memory/computational complexity relatively low. Our model processes inputs structured as $[5,1]$ or $[5,8]$ tensors, where `$5$' represents the number of neighboring spectra considered, `$1$' denotes the chromatogram, and `$8$' reflects the additional computed features. The outputs of the dilated CNN are encoded into tensors of size $[hidden\_size]$, where `$hidden\_size$' indicates the dimension of the feature space. This encoding captures information about the temporal dynamics of a specific m/z value.

The time-series embeddings are input directly into the Transformer spectrum encoder, where the self-attention mechanism learns to identify similarities in the time-series data. Ideally, ions from the same peptide will exhibit similar chromatograms, whereas ions from coeluting peptides will display different patterns. By explicitly modeling these similarities, the model's ability to differentiate between ions from coeluting peptides is improved.

In addition, we adopt the FlashAttention 2 \cite{dao2023flashattention} implementation of the attention function, allowing us to process very large DIA spectra, with high computation and memory efficiency.

\subsection*{RoPE Integration}
GraphNovo employs the Graphformer\cite{yang2021graphformers} graph neural network to encode the spectrum graph, capturing comprehensive information about the spectrum. However, applying this approach to DIA data is impractical due to the large size of DIA spectra, which results from high levels of coelution. To address this challenge, we replace the memory-intensive node and path encoders used in Graphformer with Rotary Position Embedding (RoPE). This step is also neccessary for the adoption of Flash Attention 2, since it does not allow us to operate directly on the attention matrix, which is required by Graphformer.

RoPE is a position encoding technique that introduces rotational invariance by representing positional information in a circular format\cite{su2024roformer}. In our approach, RoPE encodes the mass differences between graph nodes, allowing the model to learn meaningful mass relationships automatically. This effectively transforms the spectrum graph into a fully-connected graph where edges represent mass differences, enabling the model to identify the most relevant mass differences for the task without the computational overhead of traditional node and path encoders. The adoption of RoPE is advantageous, as it captures the critical information conveyed by the mass difference between pairs of graph nodes, effectively representing the cumulative mass of the amino acids that lie between them.

\subsection*{Two Stage Decoder}

We adopt a two-stage decoding process, similar to GraphNovo\cite{mao2023graphnovo} to alleviate the sequence memorization issue during training, where the decoder simply remembers seen training sequences, and fail to generalize on unseen sequences. This approach includes:

\begin{itemize}
\item Stage One: Optimal Path Prediction

In the first stage, the model predicts the optimal path through the spectrum graph. This involves identifying the most probable sequence of nodes (representing graph node mass values) that form a potential peptide sequence. The graph is constructed such that each node represents a possible peptide fragment, and edges denote feasible transitions based on mass differences.
\item Stage Two: Sequence Filling

In the second stage, the optimal path is refined to generate the final peptide sequence. This stage involves filling in the mass tags along the predicted path with corresponding amino acids, ensuring the sequence adheres to known peptide fragmentation patterns.
\end{itemize}

\subsection*{Coelution-aware Pretraining}
A key difference between DDA and DIA data is that, in DDA, since precurosr selection is performed, there is typically a small number of peptides in the MS2 spectrum. Meanwhile in DIA, since we uniformly select a precursor range to perform fragmentation, there is a significant amount of coelution, i.e., one spectrum consisting of fragment ions of multiple coeluting peptides. In our study, coeluing peptides are identified by searching through the database search
result, and include those whose RT (retention time) overlaps with the target de novo peptide, and their precursor m/z falls in the same isolation window as the target peptide.

In a normal de novo sequencing algorithm, we typically only consider the fragment ions of the de novo target peptide. For instance, in the optimal path task of our program, we predict only the graph nodes (n-terminal masses of peaks) belonging to the de novo target peptide, and the next step we replace these graph nodes with amino acids, resulting in predicted peptide sequence. However, if we incorporate the information about the fragment ions of other coeluting peptides, we might be able to provide the model broader information about the spectrum, improving de novo performance. More specifically, in traditional de novo algorithms, the fragment ions of other coeluting peptides are treated as noises, but we can give them labels for the model to learn. With such information, the model can make less mistakes distinguishing noise peaks from signal peaks (of the target de novo peptide), since the model knows some noise peak is probably a fragment ion of other coeluting peptide, thus less likely predict it as a signal peak.

To achieve this, we introduce coelution-aware pretraining to our algorithm. We adopt the same Transformer encoder in Section \ref{sec:spectrum_encoder}, without the spectrum graph conversion, i.e. we keep all the original m/z values as the input to the Transformer. After layers of self-attention, we obtain the embedding for each m/z value in the spectrum.
These embeddings are trained under the ion type loss, a cross entropy loss, representing the type of fragment ions. 12 types of ions are considered (see Supporting Table \ref{table:ion_pretrain}), plus noise. These labels corresponds to the fragment ion types of all coeluting peptides, not only the target peptide. After pretraining, the trained embeddings are fed to downstream optimal path and sequence generation models as features for the Transformer encoder.

\section{Theoretical Analysis}
\label{sec:supporting_theory}
We implemented simulations to generate synthetic mass spectra reflecting the characteristic signal and noise levels associated with different acquisition methods and datasets.

\subsection*{Addition of Signal and Noise Peaks} 

Signal peaks were uniformly sampled from the theoretical peaks of the target peptide to represent peptide evidence. To simulate experimental conditions, random noise peaks were added to each spectrum, with the quantity adjusted to reflect typical noise levels for each acquisition method, introduced by different level of coelution. Specifically, for DDA, we included 50 signal peaks and 500 noise peaks; for older-generation DIA, 60 signal peaks and 1,750 noise peaks were added; and for Astral DIA, 150 signal peaks and 9,000 noise peaks were included. Noise peaks were assigned random m/z values with uniform intensities to mimic realistic spectral conditions.

For real world test dataset, we substitute the experimental parameters with real data statistics, and take into consideration the median peptide length, as well as median signal and noise intensities.

\subsection*{Calculation of XCorr Scores}

We calculated the XCorr scores for the simulated spectra by cross-correlating the experimental spectra with the theoretical spectra of the target peptides. This involved binning the spectra, normalizing the intensities, and computing the dot product between the experimental and theoretical spectra after shifting the theoretical spectrum over a range of lags to find the maximum correlation.

Computation of p-values: We employed the dynamic programming approach described by Noble, et al. \cite{noble2012pvalue} to compute the p-values associated with the XCorr scores. This method estimates the distribution of XCorr scores under the null hypothesis and determines the statistical significance of the observed XCorr scores.

\subsection*{Adjustment for Multiple Hypotheses}

Theoretically, the p-value is inversely related to peptide recall. This inverse relationship arises because, when computing the Sidak-corrected p-value \cite{sidak1967rectangular}, we must adjust for a significantly larger number of peptide hypotheses in de novo sequencing than in database searches. In database searches, only the peptides existing in the database are considered, limiting the number of comparisons. In contrast, de novo sequencing must account for all possible peptide sequences, dramatically increasing the number of hypotheses and necessitating a stricter correction for multiple testing. Consequently, the p-value threshold becomes more stringent in de novo sequencing, potentially reducing peptide recall. The relationship between peptide recall and p-value is given by the following equation:

\begin{equation}
\begin{aligned}
Peptide\ Recall &= \mathbb{E}{\left[ \frac{\# Denovo\ Success}{\# Database\ Success} \right]} \\
&\approx \frac{(1 - p)^{\#\ De\ Novo\ Peptides}}{(1 - p)^{\#\ Database\ Peptides}}
\end{aligned}
\end{equation}

where $\#\ De\ Novo\ Success$ and $\#\ Database\ Success$ are the number of successfully identified peptides by de novo mode/ database search mode, $p$ is the p value, and $\#\ De\ Novo\ Peptides$ and $\#\ Database\ Peptides$ refer to the number of peptides with the same precursor mass to compare for de novo or database search, with $\#\ De\ Novo\ Peptides \gg \#\ Database\ Peptides$

Additionally, in peptide database search, the true positive is also inversely related to the p-value, as evidenced by the probability of successful database identification computed as 

\begin{equation}
\begin{aligned}
DB\ Identification\ Prob &= \mathbb{E}{\left[ \frac{\# Database\ Success}{\# Total} \right]} \\ &\approx (1 - p)^{\#\ Database\ Peptides}
\end{aligned}
\end{equation}

Although both de novo peptide recall and database-based identification probability exhibit an inverse relationship with the corresponding p-values, the impact of increasing p-value is more pronounced in the de novo approach. This trend suggests that de novo identification methods are more sensitive to statistical confidence levels compared to database search strategies, due to their reliance on sequence inference without prior knowledge.

\section{Additional Features for Spectrum}

Below is a list of the 8 additional features about the spectrum in our model.

\begin{enumerate}
\item Normalized Mass Over Charge: We compute the mass-to-charge ratio, normalized by a predefined upper limit of $3500 Th$, to standardize this value across different spectra. Computed as $e^{\frac{m/z}{UPPER\_LIMIT}}$.
\item Relative Intensity: Each peak's intensity is expressed as a fraction of the intensity of the most abundant peak in the spectrum, facilitating comparisons across variable signal strengths. Computed as $\frac{I}{I_{max}}$, where $I_{max}$ is the intensity of most abundant peak in the spectrum.
\item Rank: Each peak is assigned a rank based on its abundance relative to other peaks, sorted from the most to the least abundant. Computed as $\frac{R}{N}$, where $R$ is the target peak's rank in terms of intensity.
\item Half Rank: This metric assigns a rank to a peak after the intensities of all peaks are halved, highlighting the relative stability of peak abundances. Computed as $\frac{R_{half}}{N}$, where $R_{half}$ is the target peak's rank with half its intensity.
\item Local Significance: Using the hyperbolic tangent function, we scale the intensity of each peak relative to the minimum intensity within the neighboring window, emphasizing peaks with significant local variance. Computed as $tanh({\frac{I}{2(I_{min}-1)}})$, where $I_{min}$ is the lowest intensity within 50 Th window.
\item Local Rank: Similar to the global rank, but restricted to peaks within the $\pm50$ Th m/z range, providing a localized perspective on peak significance. Computed as $\frac{R_{local}}{N_{local}}$, similar to rank.
\item Local Half Rank: This is computed like the half rank but limited to the local window, offering insights into the comparative dynamics of local peaks. Computed as $\frac{R_{half\_local}}{N_{local}}$.
\item Local Relative Intensity: We measure each peak's intensity relative to the most intense peak within the neighboring window, allowing for an assessment of local peak dominance. Computed as $\frac{I}{I_{local\_max}}$.
\end{enumerate}

\section{Ion Offsets}

Supporting Table \ref{table:ion_offset} includes the ion offsets for converting to n-terminal mass in our model.

\begin{supportingtable}[t!]
  \centering
\begin{tabularx}{0.4\textwidth}{ l | l  }
\hline
\textbf{Ion Type} & \textbf{Neutral Offset} \\
\hline
    \hline
    a &  [N] - CHO \\
    \hline
    b & [H] - H \\
    \hline
    y & [C] + H  \\
    \hline

\end{tabularx}
    \caption{Ion offsets for ion types we used. [N] is the molecular mass of the neutral N-terminal group; [C] is the molecular mass of the neutral C-terminal group. C,H,O are the mass of the carbon atom, hydrogen
    atom and oxygen atom individually. \cite{mao2023graphnovo}}
\label{table:ion_offset}
\end{supportingtable}

\section{Type of Labeled Ions in Pretraining}

Supporting Table \ref{table:ion_pretrain} includes the ion types considered in our pretrain model.


\begin{supportingtable}[t!]
  \centering
\begin{tabularx}{0.2\textwidth}{ l }
\hline
\textbf{Ion Type} \\
\hline
    \hline
    1a \\
    \hline
    1b \\
    \hline
    2a \\
    \hline
    2b \\
    \hline
    1a-NH$_3$ \\
    \hline
    1a-H$_{2}$O  \\
    \hline
    1b-NH$_3$ \\
    \hline
    1b-H$_{2}$O  \\
    \hline
    1y \\
    \hline
    2y \\
    \hline
    1y-NH$_3$ \\
    \hline
    1y-H$_{2}$O  \\
    \hline

\end{tabularx}
    \caption{Types of Ions Labeled during coelution-aware pretraining. For precursor charge $\leq$ 2, ions with charge 1 are considered. For precursor charge > 2, ions with charge 1 or 2 are considered.}
\label{table:ion_pretrain}
\end{supportingtable}

\section{Precursor Feature Detection}
For the detection of precursor features from liquid chromatography-mass spectrometry (LC-MS) maps, we utilized the same standard set of precursor information as employed by DeepNovo-DIA \cite{tran2019deepnovodia}. In our experiments, we implemented the detection results from DIA-NN \cite{demichev2020diann} during the main experiment, and DIAUmpire \cite{tsou2015diaumpire} during the comparison with Cascadia; however, these can be substituted with outputs from other existing peak detection algorithms, such as those referenced in Zhang et al. (2012) \cite{zhang2012peaks}, Taynova et al. (2016) \cite{tyanova2016maxquant} or Tsou et al. (2015) \cite{tsou2015diaumpire}. The outcome of this detection step is a list of precursor features, each comprising the following essential information: feature ID, precursor mass-to-charge ratio (m/z), charge state, retention-time center, and scans across the retention-time range.

Furthermore, given the m/z and retention-time range of a feature, we collected all tandem mass spectrometry (MS/MS) spectra that fell within the feature's retention-time range and whose DIA m/z windows encompassed the feature's m/z value. More specifically, our precursor information includes the following data:

\begin{itemize}

\item{Feature ID}: an unique identifier given to each precursor.
\item{Precursor m/z}: the mass-to-charge ratio of the precursor ion.
\item{Precursor Charge}: the charge state of the precursor ion.
\item{RT\_Mean}: the mean of the retention-time range.
\item{Sequence}: this column remains empty during de novo sequencing; in training mode, it contains the peptide sequences identified by the in-house database search for training purposes.
\item{Scans}: a list of all MS/MS spectra associated with the feature as described above. 
\end{itemize}

\section{Characteristics of Test Datasets}

Supporting Table \ref{tab:data_details} provides the key parameters for each test dataset, these parameters are utilized in our theoretical analysis of peptide recall.

\begin{supportingtable*}[t!]
\centering
\resizebox{\linewidth}{!}{
\begin{tabular} { l|l|l|l|l|l|l|l }
\hline

\textbf{Dataset} & \makecell[l]{\textbf{Coeluting} \\ \textbf{Number}} & 
\makecell[l]{\textbf{\# Signal} \\ \textbf{Peaks}} & 
\makecell[l]{\textbf{\# Noise} \\ \textbf{Peaks}} & 
\makecell[l]{\textbf{Median} \\ \textbf{Peptide}\\ \textbf{Length}} & 
\makecell[l]{\textbf{Median} \\ \textbf{Noise} \\ \textbf{Intensity}} & 
\makecell[l]{\textbf{Isolation} \\ \textbf{Window}} & 
\makecell[l]{\textbf{Mass} \\ \textbf{Spectrometer}} \\

\hline
\hline
Hela 5 Th & 18.50 & 72 & 1,965 & 11 & 0.44 & 5 & Q Exactive \\
Hela 10 Th & 28.03 & 80 & 1,965 & 12 & 0.52 & 10 & Q Exactive \\
Hela 20 Th & 36.20 & 91 & 2,438 & 13 & 0.56 & 20 & Q Exactive \\
PXD026600 Narrow & 5.73 & 59 & 967 & 13 & 0.64 & 8 & Fusion\\
PXD026600 Wide & 7.24 & 65 & 1,485 & 14 & 0.69 & 15 & Fusion \\
PXD026600 Overlap & 5.79 & 53 & 1,130 & 13 & 0.65 & 8 & Fusion \\
PXD026600 Variable & 5.95 & 60 & 1,275 & 13 & 0.67 & 8-15 & Fusion \\
PXD019777 & 12.34 & 91 & 3,057 & 13 & 0.73 & 24.25 & Q Exactive \\

\hline
Yeast KO & 19.49 & 152 & 9,404 & 12 & 0.34 & 3 & Astral\\
PXD046453 2p5ms & 10.15 & 152 & 12,284 & 12 & 0.21 & 2 & Astral \\
PXD046453 3p5ms & 11.99 & 155 & 14,069 & 12 & 0.18 & 2 & Astral\\
PXD046283 100ng & 4.66 & 82 & 2,790 & 12 & 0.54 & 2  & Astral \\
PXD046283 300ng & 5.01 & 105 & 4,007 & 12 & 0.49 & 2  & Astral\\
PXD046283 1000ng & 5.41 & 131 & 6,114 & 12 & 0.40 & 2  & Astral\\
PXD046283 2000ng & 6.04 & 137 & 6,960 & 12 & 0.37 & 2  & Astral\\
PXD046471 30min & 6.97 & 136 & 6,526 & 12 & 0.34 & 2  & Astral\\
PXD046471 7min & 14.04 & 130 & 6,359 & 12 & 0.51 & 4  & Astral\\

\hline
\end{tabular}}
\caption{Experimental Parameters for test datasets. Coeluting number refers to how many coeluting peptides one peptide has on average, number of signal or noise peaks refers to the median number of peaks of the neighboring five spectrums which are fragment ions or the target peptide or not, median noise intensity refers to the ratio between median intensity for noise peaks and signal peaks, and isolation window has unit Th.}
\label{tab:data_details}
\end{supportingtable*}

\section{Links to Datasets}

Supporting Table \ref{tab:data_link} provides links to our training and testing datasets. Every dataset is searched with DIA-NN 1.8.1 \cite{demichev2020diann} with carboxymethylation of cysteine as fixed modification, and oxidation of methionine as variable. PSMs under 1\% FDR is kept for training or testing. The raw files are converted to mzML with msConvert 3.0 with vendor peak picking. Then the PSMs and mzMLs are converted to our customized format with our processing codes, by extracting the related spectrums of each PSM. For de novo only analysis, feature detection is done on the mzML files with DIA-Umpire 2.2.8 \cite{tsou2015diaumpire}.

\begin{supportingtable}[t!]
\centering
\resizebox{\linewidth}{!}{
\begin{tabular} { l|l }
\hline
\textbf{Dataset} &
\textbf{Link} \\
\hline
\hline
Hela \cite{ting2017pecan} & Chorus Project, number 1105\\
Pain \cite{muntel2015ocuti} & \url{ftp://PASS00706:YP9554a@ftp.peptideatlas.org/}\\
PXD003179 \cite{tsou2016untargeted} & \url{https://proteomecentral.proteomexchange.org/cgi/GetDataset?ID=PXD003179} \\
PXD026600 \cite{gotti26600} & \url{https://proteomecentral.proteomexchange.org/cgi/GetDataset?ID=PXD026600} \\
PXD019777 \cite{kalxdorf19777} & \url{https://proteomecentral.proteomexchange.org/cgi/GetDataset?ID=PXD019777} \\

\hline
PXD046386 \cite{guzman2024astraldia} & \url{https://proteomecentral.proteomexchange.org/cgi/GetDataset?ID=PXD046386}\\
PXD046453 \cite{guzman2024astraldia} & \url{https://proteomecentral.proteomexchange.org/cgi/GetDataset?ID=PXD046453} \\
PXD046444 \cite{guzman2024astraldia} & \url{https://proteomecentral.proteomexchange.org/cgi/GetDataset?ID=PXD046444} \\
PXD046283 \cite{guzman2024astraldia} & \url{https://proteomecentral.proteomexchange.org/cgi/GetDataset?ID=PXD046283} \\
PXD046471 \cite{guzman2024astraldia} & \url{https://proteomecentral.proteomexchange.org/cgi/GetDataset?ID=PXD046471} \\

\hline
\end{tabular}}
\caption{Links to each dataset}
\label{tab:data_link}
\end{supportingtable}

\section{Model Implementation Details}

Our model is trained on a Lion optimizer \cite{chen2023lion} with learning rate $2\times 10^{-6}$, over 3 epochs, with early stop based on validation error. The model includes 4 layers both on encoder and decoder side, with 1,024 hidden size and 8 attention heads.

For older-generation data, training and validation sets are Pain, PXD019777, and PXD003179, with 680,947/254,467/1,206,052 PSMs respectively. For Astral data, they are PXD046386, PXD046453, and PXD046444, with 1,896,662, 1,086,577, and 2,136,215 PSMs respectively, these PSMs are chosen from the DIA-NN search result with 1\% FDR.

Each training scheme was associated with a held-out validation set comprising 20,000 peptide-spectrum matches (PSMs). These validation examples were used strictly for hyperparameter tuning and early stopping. No validation data was used in model selection or test-time inference.

For the test set, we ensured strict non-overlap in both peptide sequences and experimental conditions. All test sequences were drawn from datasets entirely distinct from those used in training. For example, when training on the human datasets PXD046453 + PXD046444, we tested only on the yeast dataset PXD046386. Conversely, when training on the yeast dataset PXD046386, we used entirely separate human datasets for testing.

\section{Sensitivity to Coelution Number}

In this section, we investigate how our algorithm responds to varying levels of coelution. The coelution number is defined as the number of peptides that coelute with a target peptide, specifically when their precursor m/z values fall within the same precursor isolation window and their retention times overlap.

To visualize the relationship between coelution number and de novo sequencing performance, we generated a plot (shown in Supporting Figure \ref{fig:386coelution}) that demonstrates our algorithm's performance across different levels of coelution. The plot illustrates that our algorithm maintains consistent performance irrespective of the coelution number, indicating its robustness in handling complex spectral data. 

\begin{supportingfigure}[t!] 
\centering \includegraphics[width=0.45\textwidth]{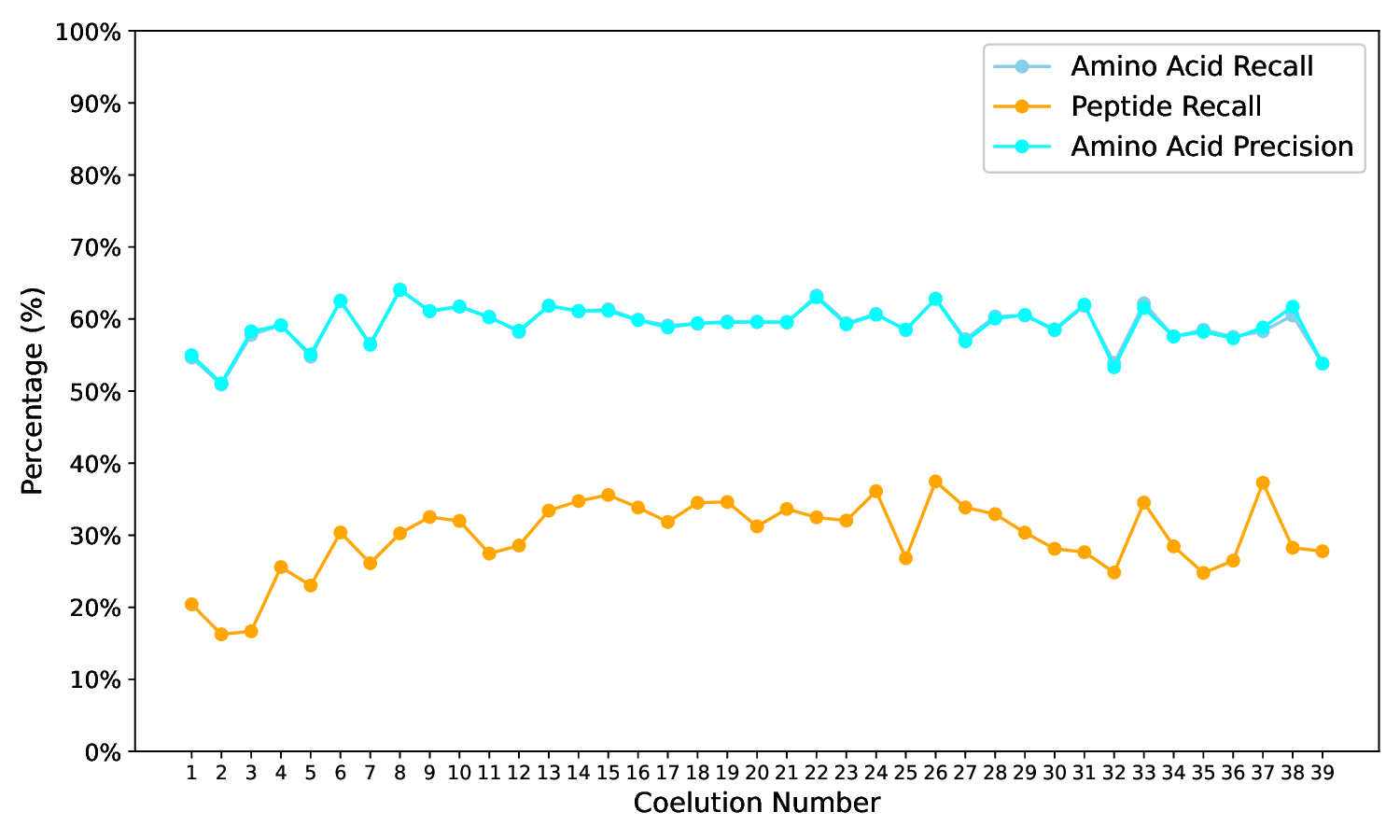} 
\caption{Performance of our method vs baselines, on the Yeast KO Dataset, on peptides with varying coelution number.} 
\label{fig:386coelution} 
\end{supportingfigure}

The ability of our method to maintain performance in the presence of numerous coeluting peptides underscores its effectiveness in dealing with the inherent complexity of DIA data. This resilience is critical for practical applications in proteomics, where accurate peptide detection amidst complex mixtures is essential.

\section{Ablation Study}

In this ablation study, illustrated in Supporting Figure \ref{fig:ablation}, we evaluate the impact of different configurations on our model using the Yeast KO dataset. The removal of coelution-aware pretraining reduces peptide recall by 8.5\% compared to DIANovo. The removal of Rotary Positional Encoding (RoPE) results in peptide recall decreasing to 67.2\% of the original. Excluding extended peak features or limiting the input to only three neighboring spectra leads to moderate decreases in performance. Finally, binning spectra into 0.5 Th intervals produces a substantial reduction in performance.

\begin{supportingfigure*}[t!]
\centering
\hfill
\begin{subfigure}[b]{0.28\textwidth}
   \includegraphics[width=\textwidth]{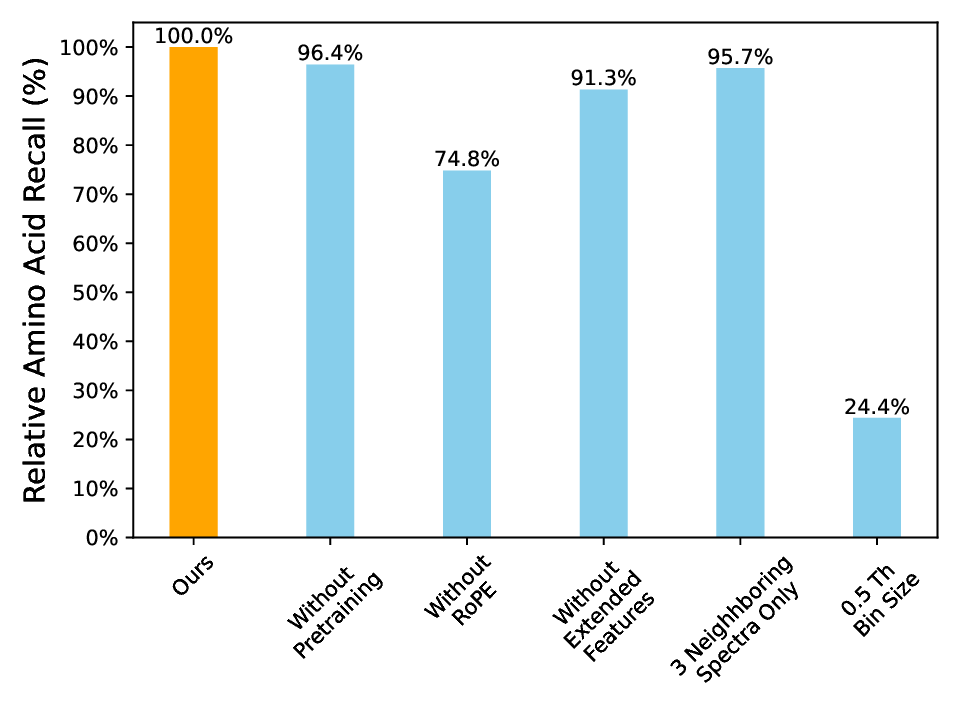}
   \caption{Relative amino acid recall}
   \label{fig:aa_ablation} 
\end{subfigure}
\hfill
\begin{subfigure}[b]{0.28\textwidth}
   \includegraphics[width=\textwidth]{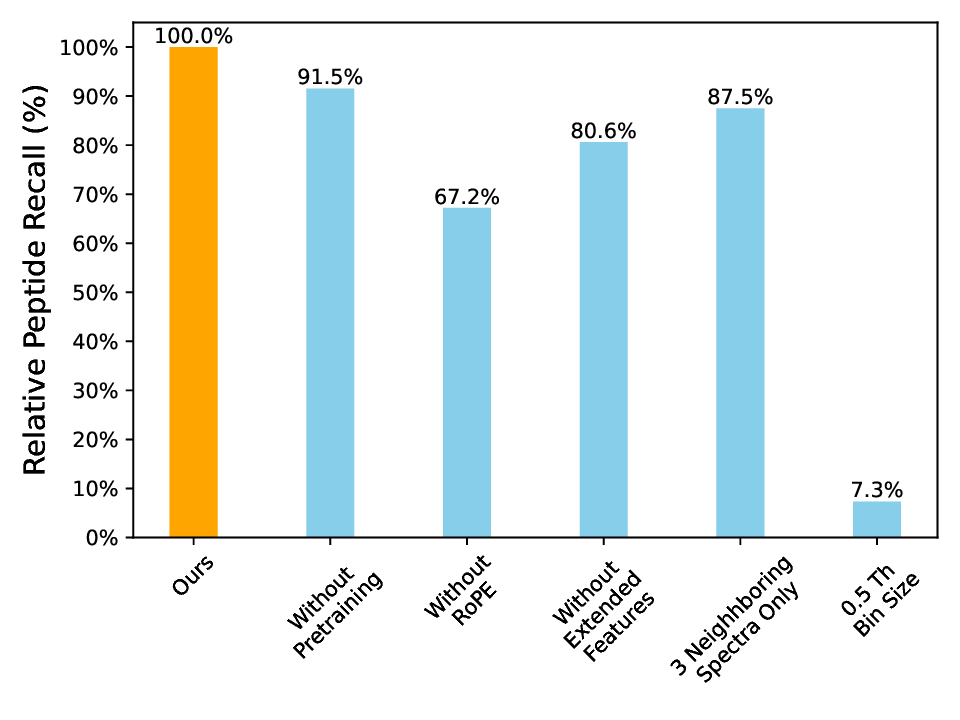}
   \caption{Relative peptide recall}
   \label{fig:pep_ablation}
\end{subfigure}
\hfill
\begin{subfigure}[b]{0.28\textwidth}
   \includegraphics[width=\textwidth]{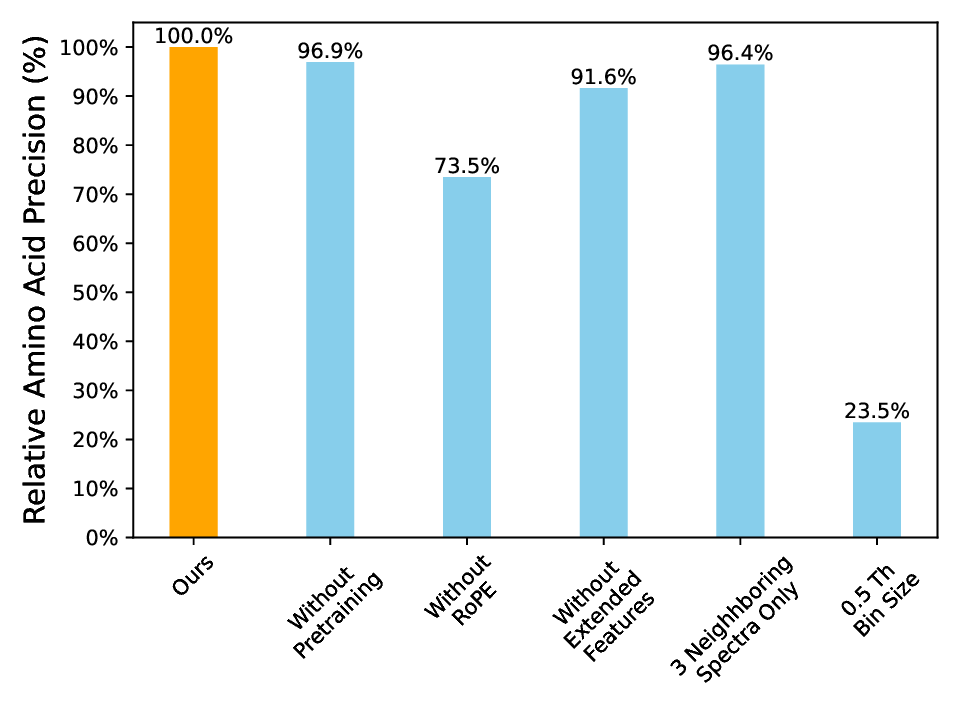}
   \caption{Relative amino acid precision}
   \label{fig:precision_ablation}
\end{subfigure}
\hfill
\caption{Relative amino acid recall (a), peptide recall (b), and amino acid precision (c) of our method vs different configurations, compared to our method, on the Yeast KO dataset.}
\label{fig:ablation}
\end{supportingfigure*}

\section{Visualization of Learned Embeddings}

In this section, we present a visualization of the learned peak embeddings to better understand the features captured by our pretrained model.

We generate a t-SNE \cite{van2008tsne} plot from the learned peak embeddings, and observe that the model implicitly captures the relationship between coeluting precursors and their corresponding fragment ions, even when trained solely with the ion type loss. The t-SNE plots (shown in Supporting Figure \ref{fig:tsne1}, \ref{fig:tsne2}, and \ref{fig:tsne3}) further reveal that peak embeddings are organized not only by fragment ion type (ion type label) but also by their source peptide (ion source label). For instance, in Supporting Figure \ref{fig:tsne1}, a dark blue cluster on the right side of the graph represents peaks originating from peptide 1 in the ion source plot. Within this cluster, the peaks are further subdivided into distinct colors (pink and green) in the ion type plot, corresponding to different fragment ion types. We provide the t-SNE plots from three different peptides to illustrate this effect.

These results suggest that the model inherently learns to associate each fragment ion with its coeluting peptide, highlighting its ability to extract meaningful structural relationships from the data.

\begin{supportingfigure*}[t!]
\centering
\hfill
\begin{subfigure}[b]{0.4\textwidth}
   \includegraphics[width=\textwidth]{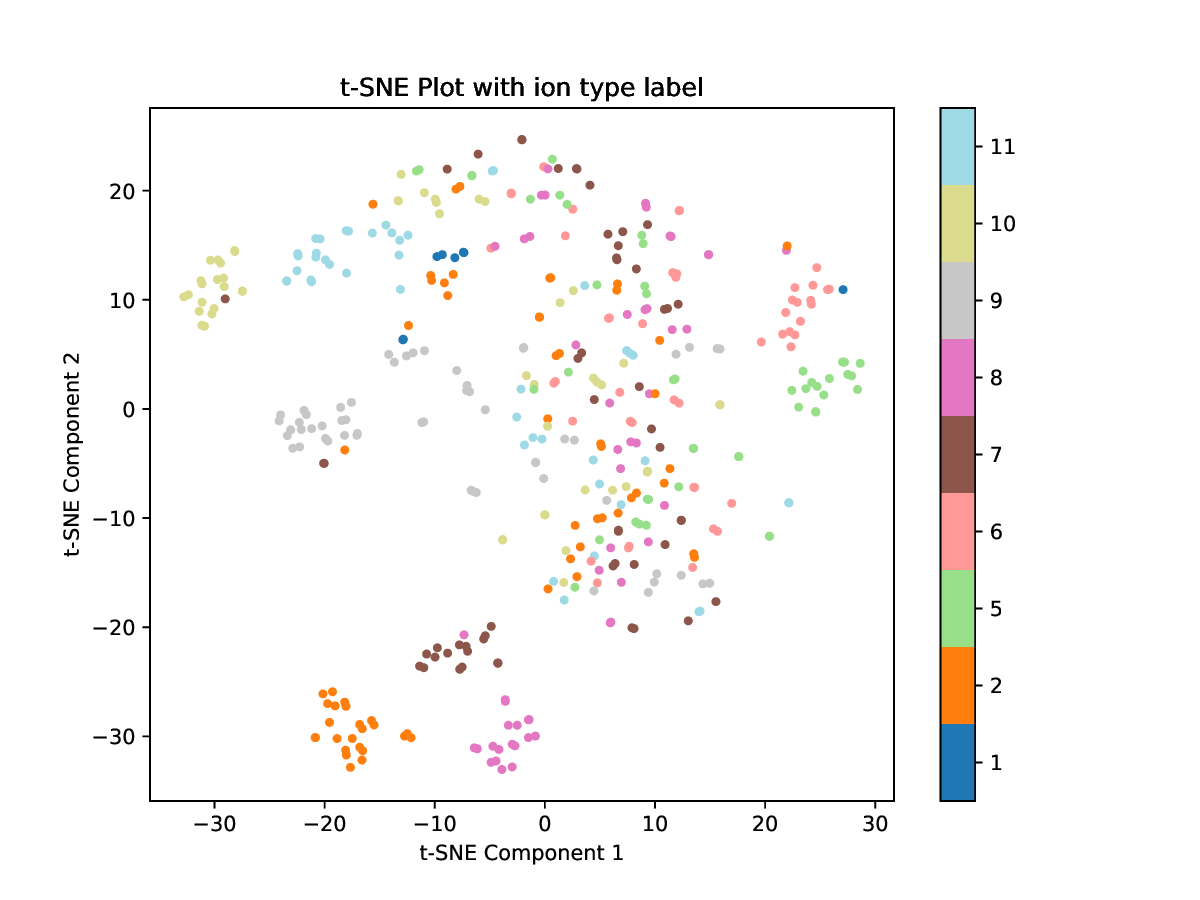}
   \caption{Ion Type Label}
   \label{fig:tsne1_iontype} 
\end{subfigure}
\hfill
\begin{subfigure}[b]{0.4\textwidth}
   \includegraphics[width=\textwidth]{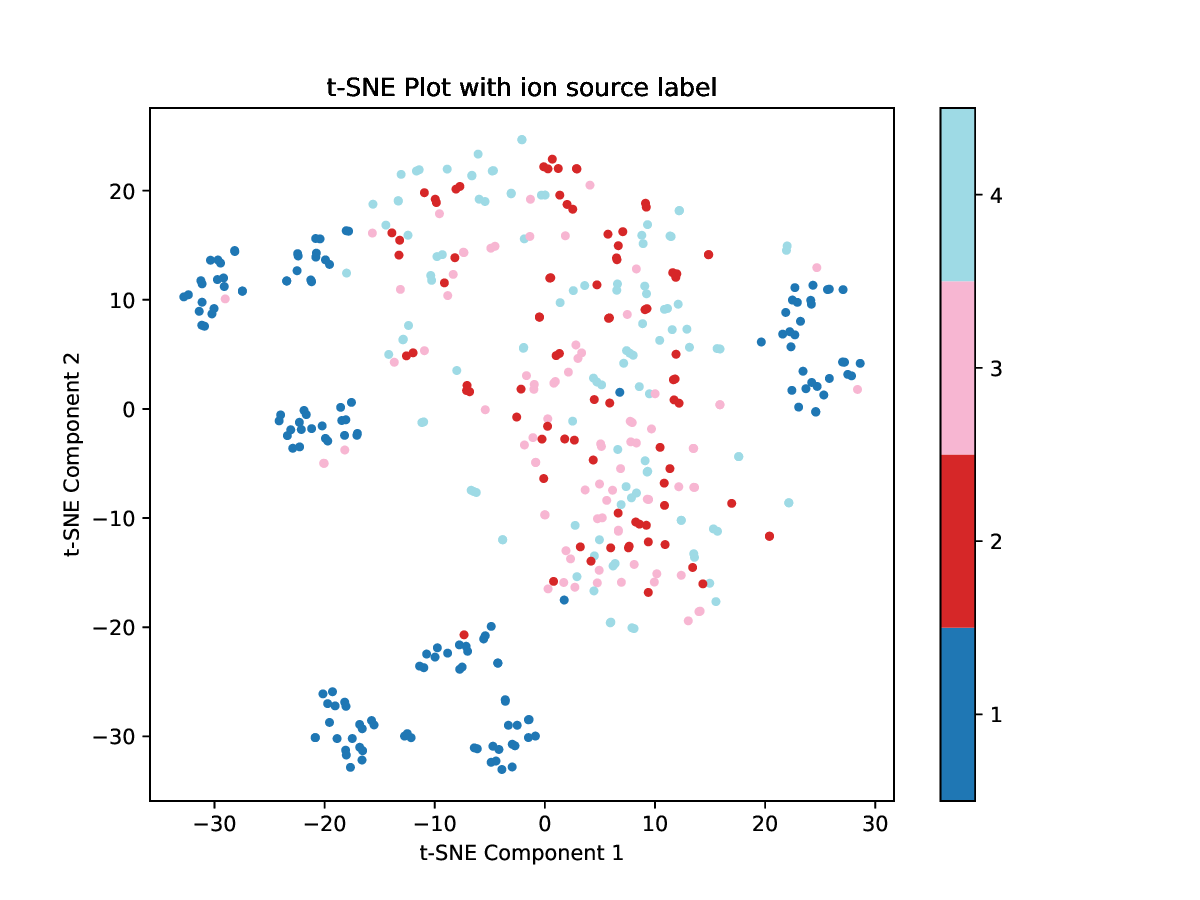}
   \caption{Ion Source Label}
   \label{fig:tsne1_ionsource}
\end{subfigure}
\hfill
\caption{tSNE plot for ion type label (a) and ion source label (b) of peptide DHGEGGIIVGSALENK2. The color for ion type label refers to the fragment ion type of each peak, while the color for ion source label refers to which coeluting peptide a peak comes from. Noise peaks (which do not belong to any coeluting peptide) are excluded.}
\label{fig:tsne1}
\end{supportingfigure*}

\begin{supportingfigure*}[t!]
\centering
\hfill
\begin{subfigure}[b]{0.4\textwidth}
   \includegraphics[width=\textwidth]{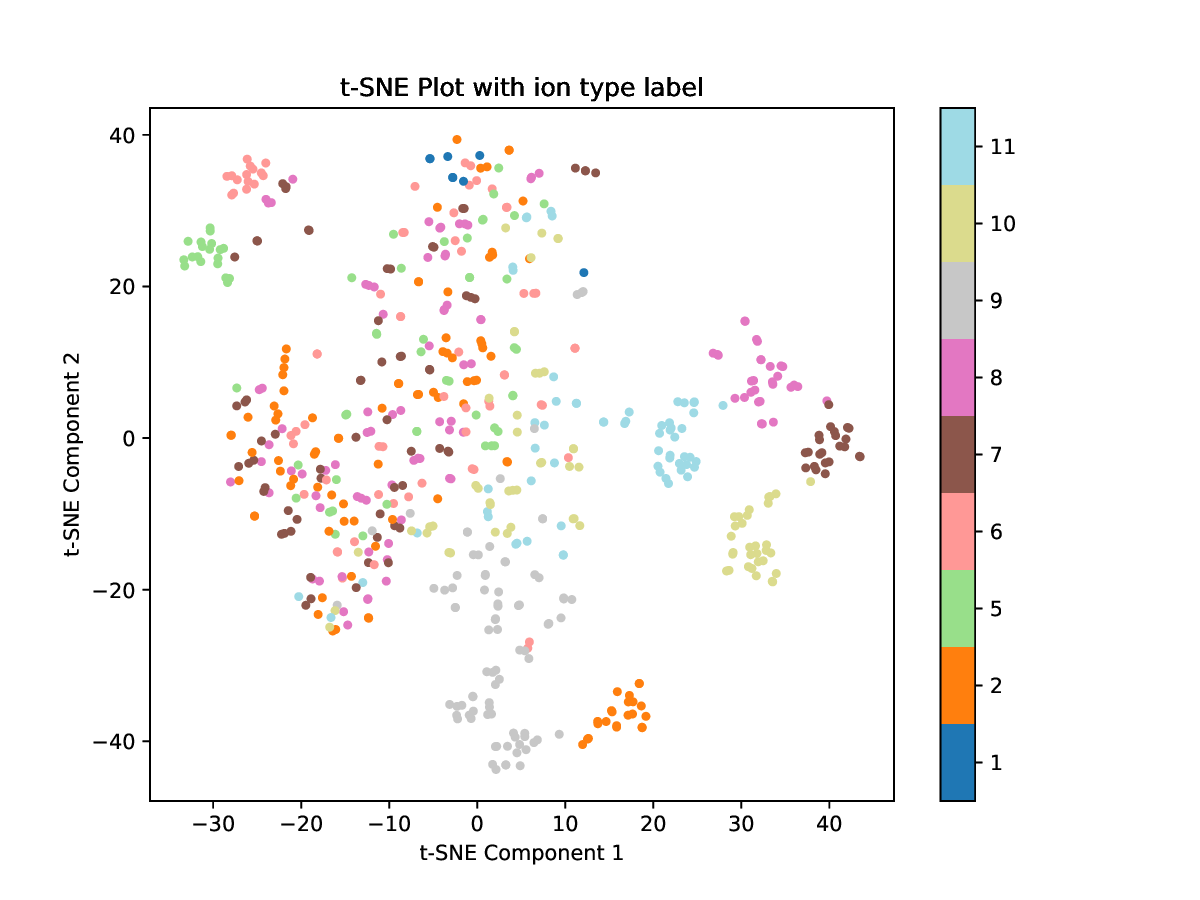}
   \caption{Ion Type Label}
   \label{fig:tsne2_iontype} 
\end{subfigure}
\hfill
\begin{subfigure}[b]{0.4\textwidth}
   \includegraphics[width=\textwidth]{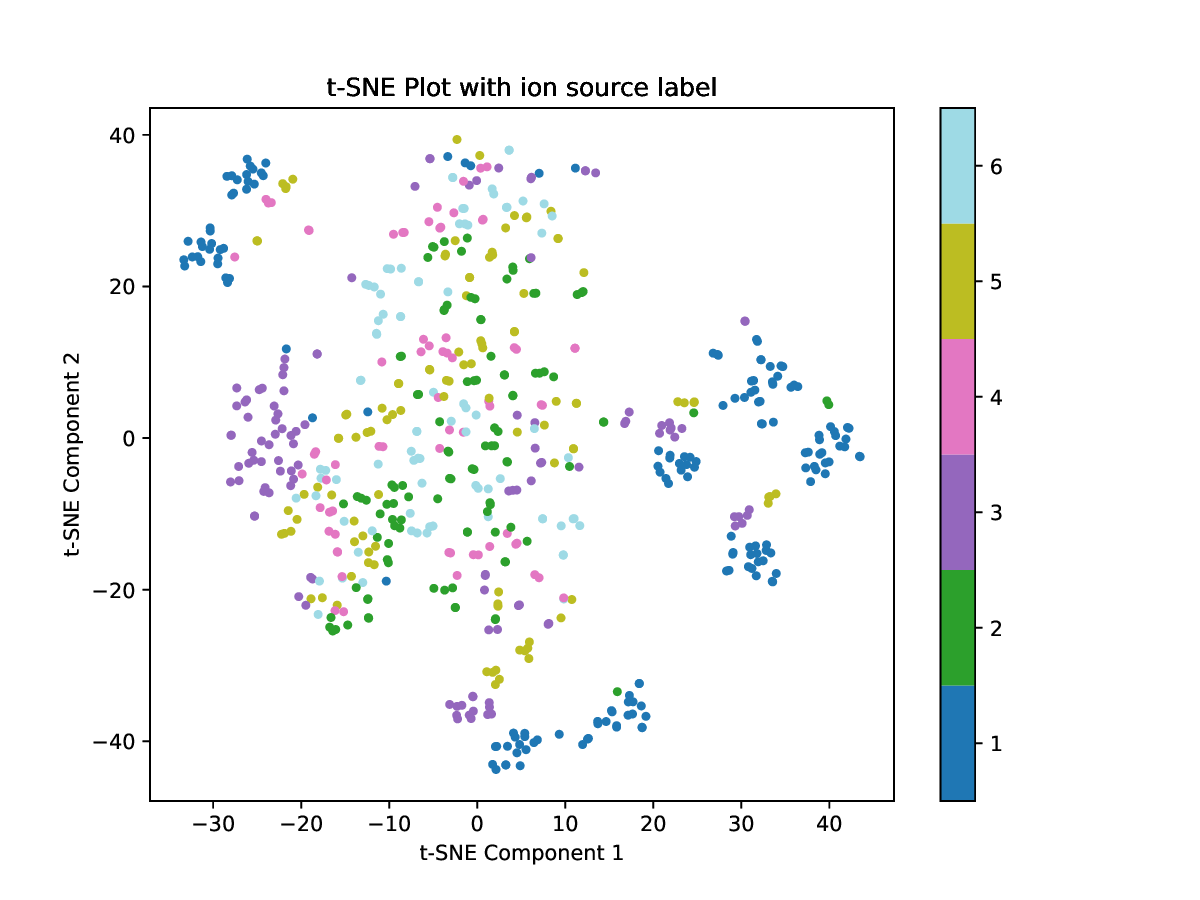}
   \caption{Ion Source Label}
   \label{fig:tsne2_ionsource}
\end{subfigure}
\hfill
\caption{tSNE plot for ion type label (a) and ion source label (b) of peptide GYWGTNLGQPHSLATK2.}
\label{fig:tsne2}
\end{supportingfigure*}

\begin{supportingfigure*}[t!]
\centering
\hfill
\begin{subfigure}[b]{0.4\textwidth}
   \includegraphics[width=\textwidth]{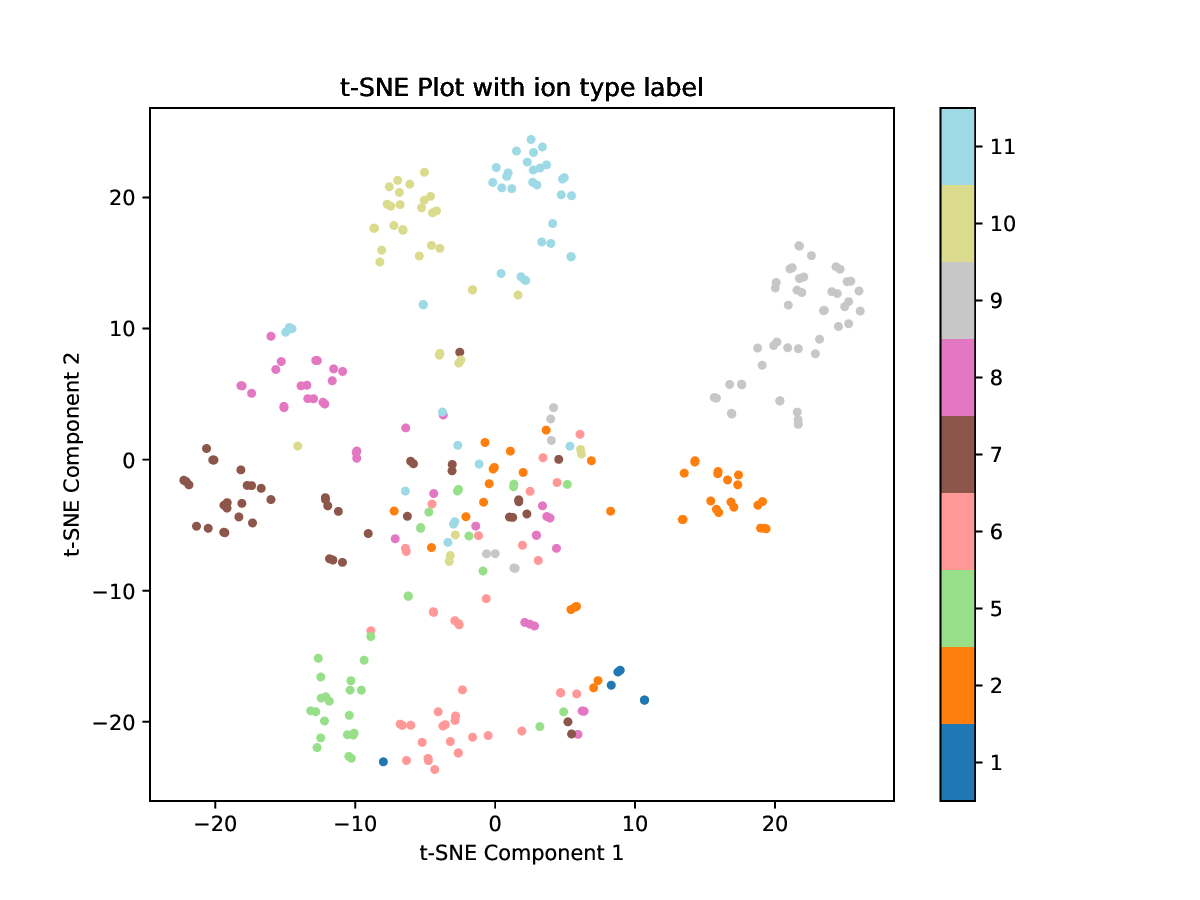}
   \caption{Ion Type Label}
   \label{fig:tsne3_iontype} 
\end{subfigure}
\hfill
\begin{subfigure}[b]{0.4\textwidth}
   \includegraphics[width=\textwidth]{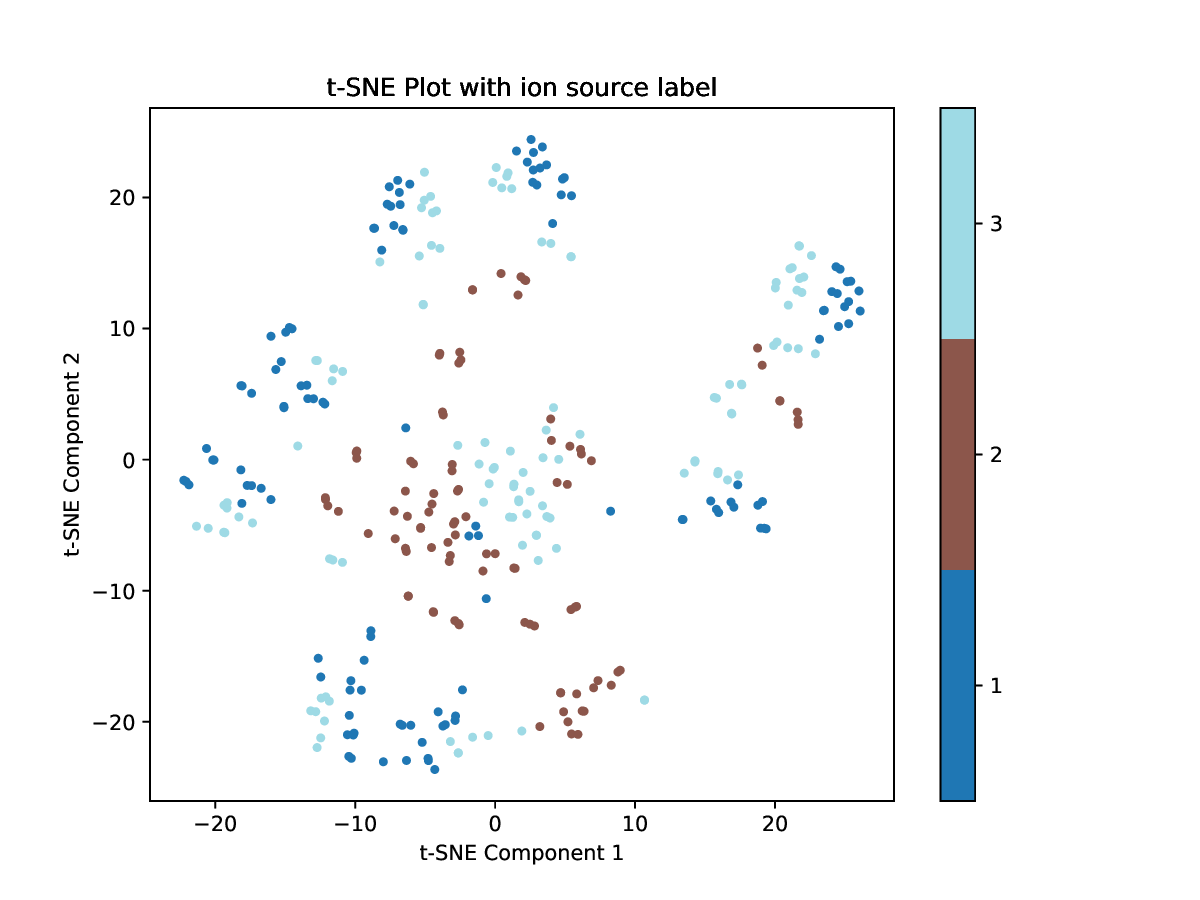}
   \caption{Ion Source Label}
   \label{fig:tsne3_ionsource}
\end{subfigure}
\hfill
\caption{tSNE plot for ion type label (a) and ion source label (b) of peptide EYLPEMAASYSHPK2.}
\label{fig:tsne3}
\end{supportingfigure*}
\end{appendices}
\end{document}